\documentclass[longauth]{aa}  

\usepackage{graphicx}
\usepackage{txfonts}
\usepackage{booktabs}
\usepackage{hyperref}

\hypersetup{
    colorlinks=true,
    citecolor=blue,
}
\usepackage{xcolor}

\begin{document} 

\title{Disk Evolution Study Through Imaging of Nearby Young Stars (DESTINYS): PDS\,111, an old T\,Tauri star with a young-looking disk}

\author{Annelotte Derkink
\inst{\ref{inst_ams}}
\and
    Christian Ginski \inst{\ref{inst_galway}}
    \and
    Paola Pinilla \inst{\ref{inst_UCL}}
    \and
    Nicolas Kurtovic \inst{\ref{inst_mpie}}
    \and
    Lex Kaper \inst{\ref{inst_ams}}
    \and
    Alex de Koter \inst{\ref{inst_ams}, \ref{inst_leuven}}
    \and
    Per-Gunnar Valegård \inst{\ref{inst_ams}}
    \and
    Eric Mamajek \inst{\ref{inst_jet}, \ref{inst_roch}}
    \and
    Frank Backs \inst{\ref{inst_ams}}
    \and
    Myriam Benisty \inst{\ref{inst_nice},\ref{inst_grenoble}}
    \and
    Til Birnstiel \inst{\ref{inst_UOmun}}
    \and
    Gabriele Columba \inst{\ref{inst_padua}}
    \and
    Carsten Dominik \inst{\ref{inst_ams}}
    \and
    Antonio Garufi \inst{\ref{inst_inaf}}
    \and
    Michiel Hogerheijde \inst{\ref{inst_ams},\ref{inst_leiden}}
    \and
    Rob van Holstein \inst{\ref{inst_esochili}}
    \and
    Jane Huang \inst{\ref{inst_michigan}}
    \and
    François Ménard \inst{\ref{inst_grenoble}}
    \and
    Christian Rab \inst{\ref{inst_UOmun},\ref{inst_mpie}}
    \and
    María Claudia Ramírez-Tannus \inst{\ref{inst_mpia}}
    \and
    Álvaro Ribas \inst{\ref{inst_camb}}
    \and 
    Jonathan P. Williams \inst{\ref{inst_hawaii}}
    \and 
    Alice Zurlo \inst{\ref{inst_chilieii}, \ref{inst_chiliyems}}
}

\institute{Anton Pannekoek Institute for Astronomy (API), University of Amsterdam, Science Park 904, 1098 XH Amsterdam,
The Netherlands\label{inst_ams},
              \email{a.r.derkink@uva.nl}
              \and
              School of Natural Sciences, University of Galway, University Road, H91 TK33 Galway, Ireland \label{inst_galway}
              \and
              Mullard Space Science Laboratory, University College London, Holmbury St Mary, Dorking, Surrey RH5 6NT, UK \label{inst_UCL}
                     \and
              Max-Planck-Institut für extraterrestrische Physik, Giessenbachstrasse 1, 85748 Garching, Germany \label{inst_mpie}
                \and
   	    Instituut voor Sterrenkunde, KU Leuven, Celestijnenlaan 200D bus 2401, 3001 Leuven, Belgium \label{inst_leuven}
                \and
              Jet Propulsion Laboratory, California Institute of Technology, Pasadena, CA, USA \label{inst_jet}
              \and 
               Department of Physics and Astronomy, University of Rochester, Rochester, NY, USA \label{inst_roch}
            \and
            Université Côte d’Azur, Observatoire de la Côte d’Azur, CNRS, Laboratoire Lagrange, Bd de l’Observatoire, CS 34229, 06304 Nice cedex 4, France \label{inst_nice}
            \and
            Université Grenoble Alpes, CNRS, Institut de Planétologie et d’Astrophysique (IPAG), F-38000 Grenoble, France \label{inst_grenoble}
              \and
              University Observatory, Faculty of Physics, Ludwig-Maximilians-Universität, Scheinerstr. 1, D-81679 Munich, Germany \label{inst_UOmun}
              \and
              Department of Physics and Astronomy “Galileo Galilei” (DFA), University of Padua, via Marzolo 8, 35131 Padua, Italy \label{inst_padua}
              \and
              INAF, Osservatorio Astrofisico di Arcetri, Largo Enrico Fermi 5, I-50125 Firenze, Italy \label{inst_inaf}
              \and 
              Leiden Observatory, Leiden University, 2300 RA Leiden, The Netherlands \label{inst_leiden}
              \and
              European Southern Observatory, Alonso de Córdova 3107, Casilla 19001, Vitacura, Santiago, Chile \label{inst_esochili}
              \and
              Department of Astronomy, University of Michigan, 323 West Hall, 1085 S. University Avenue, Ann Arbor, MI 48109, USA \label{inst_michigan}
            \and
              Max Planck Institute for Astronomy, K\"{o}nigstuhl 17, 69117, Heidelberg, Germany \label{inst_mpia}
              \and
              Institute of Astronomy, University of Cambridge, Madingley Road, Cambridge CB3 0HA, UK \label{inst_camb}
              \and
              Institute for Astronomy, University of Hawai’i at Manoa, Honolulu, HI 96822, USA \label{inst_hawaii}
              \and
              Instituto de Estudios Astrof\'isicos, Facultad de Ingenier\'ia y Ciencias, Universidad Diego Portales, Av. Ej\'ercito Libertador 441, Santiago, Chile \label{inst_chilieii}
              \and
              Millennium Nucleus on Young Exoplanets and their Moons (YEMS), Chile \label{inst_chiliyems}
            }
        
\titlerunning{PDS\,111, an old T\,Tauri star with a young-looking disk} 
\date{}
 
  \abstract
   {The interplay between T\,Tauri stars and their circumstellar disks, and how this impacts the onset of planet formation has yet to be established. In the last years, major progress has been made using instrumentation that probes the dust structure in the mid-plane and at the surface of protoplanetary disks. Observations show a great variety of disk shapes and substructures that are crucial for understanding planet formation.}
   {We studied a seemingly old T\,Tauri star, PDS\,111, and its disk. We combined complementary observations of the stellar atmosphere, the circumstellar hot gas, the surface of the disk, and the mid-plane structure.}
   {We analyzed optical, infrared, and sub-millimeter observations obtained with VLT/X-shooter, Mercator/HERMES, TESS, VLT/SPHERE, and ALMA, providing a new view on PDS\,111 and its protoplanetary disk. The multi-epoch spectroscopy yields photospheric lines to classify the star and to update its stellar parameters, and emission lines to study variability in the hot inner disk and to determine the mass-accretion rate. The SPHERE and ALMA observations are used to characterize the dust distribution of the small and large grains, respectively.}
   {PDS\,111 is a weak-line T\,Tauri star with spectral type G2, exhibits strong H$\alpha$ variability and with a low mass-accretion rate of $1-5\times10^{-10}$\,M$_{\odot}$\,yr$^{-1}$. We measured an age of the system of 15.9$^{+1.7}_{-3.7}$\,Myr using pre-main sequence tracks. The SPHERE observations show a strongly flaring disk with an asymmetric substructure. The ALMA observations reveal a 30\,au cavity in the dust continuum emission with a low contrast asymmetry in the South-West of the disk and a dust disk mass of 45.8\,$M_\oplus$ or $\sim0.14$\,$M_{\rm{Jup}}$. The $^{12}$CO observations do not show a cavity and the $^{12}$CO radial extension is at least three times larger than that of the dust emission.}
   {Although the measured age is younger than often suggested in literature, PDS\,111 seems relatively old; this provides insight into disk properties at an advanced stage of pre-main sequence evolution. The characteristics of this disk are very similar to its younger counterparts: strongly flaring, an average disk mass, a typical radial extent of the disk gas and dust, and the presence of common substructures. This suggests that disk evolution has not significantly changed the disk properties. These results show similarities with the "Peter Pan disks" around M-dwarfs, that "refuse to evolve". 
   }
   \maketitle
%
\section{Introduction}

Disk evolution seems inherently connected to the formation of planets in circumstellar disks around low-mass pre-main-sequence (PMS) stars. Some of the big questions are when, where, and how planets, in all their observed varieties, exactly form. A thorough understanding of disk evolution is key to understand the onset of planet formation in these environments. 

An important aspect of the evolution of a disk is how matter is distributed, for instance how dust grains with different sizes are distributed vertically, radially and within substructures. In recent years, major progress has been made using telescopes and instruments that are able to probe the mid-plane structure and surface of the disk (e.g., ALMA and VLT/SPHERE). Extensive observations show great variety in disk shapes \citep[e.g.,][]{Avenhaus2018, law2023} and substructures in the form of cavities, rings, spiral arms, and arcs \citep[e.g.,][]{long2018, andrews2018, garufi2020, andrews2020, benisty2022, bae2022}. 

These substructures are thought to trace active sites in the disk, where planetesimals may grow and act as an indication of planet formation \citep{andrews2020}. This makes the study of these substructures important to uncover disk evolution and disk-planet interactions. For example, \cite{pinilla2018} studied transition disks with cavities and identified giant planet formation and dead zones as potential explanations for the observed properties of these disks. Another large study of disk substructures from \cite{garufi2020} with ALMA and SPHERE shows that rings and cavities detected with ALMA do not always have a similar SPHERE counterpart, demonstrating the importance of combined and complementary observations.

Disks around PMS stars have limited time to form planets since they are thought to dissipate rapidly. Large surveys of T\,Tauri stars have shown that only a few percent of the disks survive beyond an age of 10\,Myr \citep{Padgett2006, Carpenter2009, wahhaj2010}. 
However, recently so-called "Peter Pan Disks" have been discovered around M-type stars. These systems have accretion disks older than 20\,Myr that "refuse to grow up". These disks are thought to be long-lived due to a high initial disk mass with a low turbulence parameter and a lack of external photo-evaporation \citep{sicilia2011, Silverberg2020, Coleman2020}. 

In this paper, we combine SPHERE and ALMA observations with medium-resolution spectroscopy of the star and the hot inner disk in order to infer a detailed picture of the system PDS\,111. This T\,Tauri star, often classified as G-type \citep{Rojas2008}, is known to be in the foreground of the Orion star-forming region at a distance of 158\,pc \citep{Gaia2022} with a relatively high age \citep[15-42\,Myr,][]{Rojas2008, Bell2015, Launhardt2022}. However, this system seems to be one of these rare cases where the circumstellar disk has survived up until this significant age. The aim of this work is to reassess the age determination of the system and to measure its disk properties in order to understand how PDS\,111 fits in the current understanding of disk evolution, and how this relates to the long-lived disks in the "Peter Pan Disk" scenario. 

In the following, we describe the observations in Sec.~\ref{sec: Obs} and the data analysis of the photospheric light, scattered light, dust continuum and $^{12}$CO emission in Sec.~\ref{p2:sec:data}. The age of the system and the disk properties are presented in Sec.~\ref{p2:sec:age_and_disk}. We discuss the results in Sec.~\ref{sec:discussion} and summarize our findings in Sec.~\ref{sec: Sum}.

\section{Observations \label{sec: Obs}}

PDS\,111 was observed in the framework of the Disk Evolution Study Through Imaging of Nearby Stars (DESTINYS) ESO large program (\citealt{Ginski2020, Ginski2021}). Resolved high-resolution imaging observations were carried out with the extreme adaptive optics instrument VLT/SPHERE (\citealt{Beuzit2019}). Additionally, we obtained sub-mm observations with ALMA, medium-resolution spectroscopic observations of the central star with VLT/X-shooter (\citealt{Vernet2011}), high-resolution spectroscopy with Mercator/HERMES \citep{Raskin2011} and we used photometric observations from the Transiting Exoplanet Survey Satellite mission \citep[TESS][]{ricker2015}.

\subsection{Spectroscopy}

\subsubsection{X-shooter medium resolution spectroscopy}

Spectroscopic observations in the optical to near infrared (NIR) (300-2500\,nm) have been taken with X-shooter, mounted on the UT2 of the VLT, with program ID 108.2216.001. PDS\,111 was observed with X-shooter during two separate epochs on November 28, 2021 and December 7, 2021. The first observation has been taken with a poor seeing of $\sim$2.6\arcsec\ and the second under good seeing conditions of $\sim$0.7\arcsec. Both observations can be used in the study. The spectra are taken in nodding mode. The used slit widths are 0.5\arcsec\ in the UVB arm leading to a resolving power of 9700, and 0.4\arcsec\ in the VIS and NIR arms, which corresponds to a resolving power of 18400 and 11600, respectively. 

The spectroscopic data were reduced with the ESO X-shooter \texttt{pipeline 3.3.5} provided by ESO \citep{Modigliani2010}. The pipeline consists of a flat-field correction, bias subtraction, and wavelength calibration. Spectrophotomeric standards from the ESO database were used for flux calibration. The reduction of the nodding spectra was done with the \texttt{xsh\_scired\_slit\_nod} recipe provided in the software package from ESO. The \texttt{Molecfit 1.5.9} tool was used to correct for the telluric sky lines in the spectra \citep{Smette2015, Kausch2015}.

The arms are normalized separately using second-order spline functions. The continuum points for the spline fit are selected at the same wavelength intervals for each epoch. 

\subsubsection{HERMES high resolution spectroscopy}

High-resolution follow-up observations have been taken with the HERMES spectrograph (380-900\,nm) mounted on the {\it Mercator} 1.2\,m telescope on La Palma \citep{Raskin2011} to obtain a longer temporal baseline for line variability. HERMES has a spectral resolving power of R\,$\sim$85,000. Observations were taken on two nights, October 20, 2022 and October 28, 2022. The data were reduced with the standard HERMES pipeline, consisting of a flat-field correction, bias subtraction, wavelength calibration, and cosmic removal. The humidity was relatively high ($\sim$50\%) for the second observation compared to the first observation, introducing stronger telluric features. The spectra were rebinned on the same resolution as X-shooter to increase the signal-to-noise ratio and similarly normalized in order to compare the spectra of HERMES and X-shooter. 

\begin{figure*}
\centering
\includegraphics[width=0.999\textwidth]{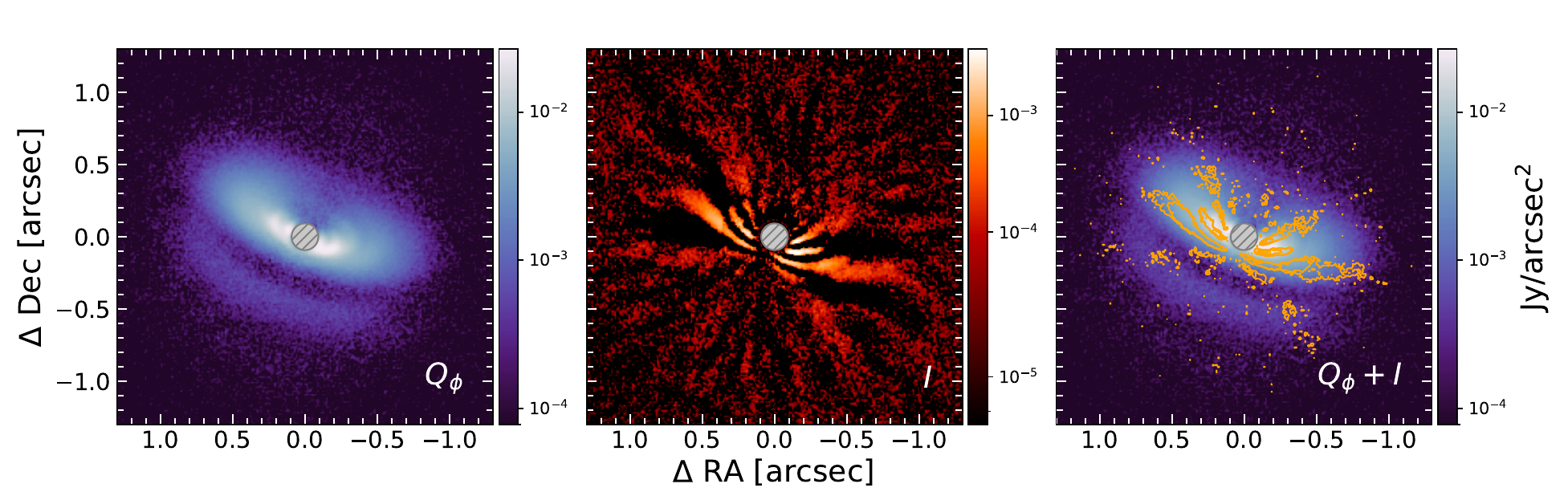}
\caption[]{SPHERE/IRDIS H-band (1.65\,$\mu$m) observations of PDS\,111. In all images the star is at the center and the images are aligned with the North-axis up and East-axis to the left. \textit{Left:} Polarized light observation. We show the azimuthal Stokes parameter $Q_\phi$ which contains all azimuthally polarized scattered light as a positive signal. 
\textit{Middle:} Same data set as in the left panel, but reduced with angular differential imaging, showing total intensity. Due to the reduction approach, the disk signal is strongly self-subtracted removing particularly low spatial frequencies. The shape of the detected high spatial frequency sub-structure is likely distorted. For the same reason, the given surface brightness is a lower limit only, which depends in a complex way on the local self subtraction in the image.
\textit{Right:} Same image as in the left panel, overlayed with contours (5\,$\sigma$, 14\,$\sigma$, 68\,$\sigma$, 140\,$\sigma$, 205\,$\sigma$; relative to the image background noise) from the center panel to highlight the structures that are traced in both images. 
In all images, the gray, hashed circle in the center marks the area that was covered by the coronagraphic mask and where the star is located.
All images are shown on a log scale to highlight the disk morphology. Appendix~\ref{app:pol-flux} shows a flux calibrated version of the polarimetric image.}
\label{fig: qphi-adi-main}
\end{figure*}

\subsection{SPHERE/IRDIS near-infrared imaging}
PDS\,111 was observed with the InfraRed Dual-band Imager and Spectrograph (IRDIS, \citealt{Dohlen2008}) subsystem of SPHERE on December 30, 2020. The observations were carried out in the BB\_H filter with a central wavelength of 1.65\,$\mu$m. We used the dual-beam polarimetric imaging mode of the instrument (\citealt{deBoer2020,vanHolstein2020}) in pupil-tracking mode \citep{holstein2017}. The central star was placed behind an apodized Lyot coronagraph with an effective radius of 92.5\,mas to avoid saturation and to limit photon noise. We recorded a total of 104 science frames split into 26 polarimetric cycles. The individual frame exposure time was 32\,s resulting in a total exposure time of 55.5\,min. The atmospheric conditions during the observation sequence were excellent with an average seeing of 0.72\arcsec\ and a coherence time of 9.8\,ms. Bracketing the science observations we acquired dedicated astrometric star-center and flux calibration frames, as well as sky calibration frames.

We reduced the data using the IRDIS Data reduction for Accurate Polarimetry\footnote{https://irdap.readthedocs.io} pipeline (IRDAP, \citealt{vanHolstein2020}), which performs polarimetric differential imaging \citep[PDI,][]{Kuhn2001} to remove the light from the bright central star. We used the default setup parameters. IRDAP first performs image pre-processing, which includes sky subtraction, flat-fielding, and bad-pixel masking. A static master flat-field image, as well as a static badpixel mask were used, which are included in IRDAP. Then the frames were separated in the left and right detector side and individually centered, using the astrometric star-center calibration frames. After these pre-processing steps IRDAP combines images from both detector sides, as well as images at different half-wave plate position to compute the Stokes $Q$ and $U$ images. To remove residual instrumental polarization a full Mueller matrix model of the telescope and instrument is implemented that describes the instrumental polarization effects. Additionally, residual (inter)stellar polarization is removed by measuring the signal at the adaptive optics correction radius within the Q and U images and subtracting a scaled total intensity image. The final $Q$ and $U$ images are then converted into the azimuthal Stokes parameters $Q_\phi$ and $U_\phi$, following the definition in \cite{deBoer2020}. The $Q_\phi$ image contains all azimuthally polarized light as positive signal. Since this is the expected orientation of the polarization direction for single scattering events it gives a good representation of the disk polarized light image. We show the $Q_\phi$ image in Fig.~\ref{fig: qphi-adi-main} and the full set of Stokes images in the Appendix (Fig.~\ref{fig: app: flux-pol}).

In addition to the PDI reduction, we performed angular differential imaging (ADI, \citealt{Marois2006}) post-processing with the IRDAP pipeline. We used the principal component analysis approach to fit the reference catalog (consisting of all science images themselves) to each science frame. For the processing the default settings of IRDAP were utilized. The ADI processed total intensity image is likewise shown in Fig.~\ref{fig: qphi-adi-main}. For this image we specifically used 4 principal components and optimization annuli between 10 and 30 pixels (0.12" to 0.37"), 30 and 100 pixels (0.37" to 1.23"), and 100 and 512 pixels (1.23" to 6.27").

\subsection{ALMA observations} \label{ALMA_obs}
The 1.25\,mm emission of PDS\,111 was observed with ALMA Band 6 as part of the ALMA project 2021.1.01705.S (PI: C.~Ginski), with 1.31\,min of exposure on August 28, 2022 and 1.31\,min in September 2, 2022. The correlator was configured to observe four spectral windows: one included the $^{12}$CO molecular line at $230.538\,$GHz in the J=2-1 transition with a frequency spacing of $488.3$\,kHz ($\sim$0.65\,km\,s$^{-1}$ per channel), while the remaining three were targeted at dust continuum emission centered at $233.519$\,GHz, $246.020$\,GHz, and $248.021$\,GHz, with a channel frequency spacing of $15.625\,$MHz and a total bandwidth of $2$\,GHz. The antenna baselines span 15.1\,m to 1210.6\,m.

Executing the \texttt{scriptforPI} provided by ALMA, we obtain the pipeline calibrated data, which is self-calibrated using dust continuum emission observations, so that the signal-to-noise ratio can be improved. Using \texttt{CASA 5.6.2} we flagged the channels located at $\pm 20\,$km\,s$^{-1}$ from the line center to obtain the dust continuum emission. A pseudo-continuum measurement set is created by combining the remaining channels with the continuum spectral windows. To reduce data volume, the set was averaged into 125\,MHz channels and `$6$s' bins. The self-calibration was applied on the continuum by imaging with the \texttt{CLEAN} algorithm, with a Briggs robust parameter of 0.5. The robust parameter is used to control the trade-off between the sensitivity and angular resolution of the data by applying a natural and/or uniform weighting scheme to the data. The value of 0.5 slightly favors angular resolution over sensitivity \citep{briggs1995, boone2013}.  We applied four phase calibrations, using solution intervals of length "inf", "40s", "18s", and "6s", while the amplitude calibration was done with a solution interval of "6s", combining spectral windows and scans to find the solution. The correction tables were later applied to the $^{12}$CO spectral window.

After self-calibration, the continuum emission was imaged with a robust parameter of -0.5 to maximize the balance between angular resolution and sensitivity. The beam size of the $^{12}$CO is 0.429\arcsec. The gas emission was imaged with a robust parameter of 0.0 and no visibility tapering to focus on best possible angular resolution. Both images were cleaned until the $4\sigma$ noise level to apply the JvM correction \citep[JvM "effect" from][]{jorsater1995}, which takes in account the volume ratio ($\epsilon$) values of the point spread function of the images and the restored Gaussian from the \texttt{CLEAN} algorithm following the method as described in \cite{czekala2021}. We found $\epsilon$ values of $0.909$ for the dust continuum image and $0.694$ for the $^{12}$CO, which means that the dirty beam shape for the dust continuum is more closely described by an elliptical Gaussian, while that for the $^{12}$CO imaging deviates more significantly.

\noindent

\subsection{TESS photometry}
TESS observed PDS\,111 during its runs starting in November and December 2018 (target name 24697724). The target was observed with exposures of 120\,s during 42 days (spread between November, December, and January) with the TESS detector bandpass that spans from 600 - 1000\,nm. 

\section{Data analysis}\label{p2:sec:data}

\subsection{Photospheric spectrum PDS\,111}\label{sec:stellarprops}
\begin{figure*}
\centering
\includegraphics[width=0.6\hsize]{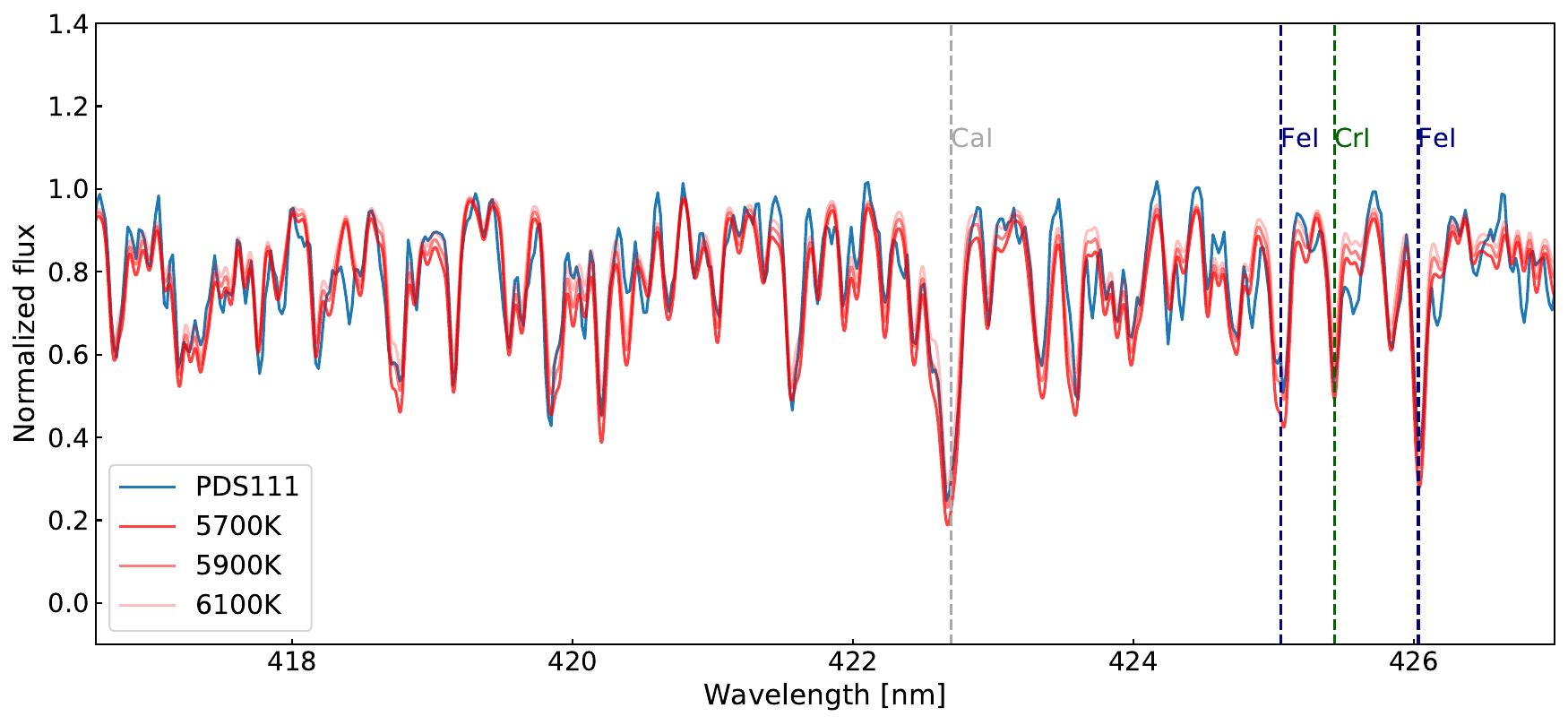}
\caption[]{The stacked spectrum of PDS\,111 is shown together with a subset of Kurucz models with different temperatures. The models have a log\,$g$ of 4.2 and a $v$\,sin$i$ of 21 km\,s$^{-1}$.}
\label{p2:fig:models}
\end{figure*}

Based on the strength of the G-band, Fe\,{\sc i} lines and H$\gamma$ around 432\,nm, the star is classified as early G-type \citep{gray2000digital}. A comparison of the spectrum with stars in the X-shooter library indicates a spectral class between G1 and G4. 
The luminosity class V is based on the relative strength of Mn\,{\sc i} 403.0\,nm and Fe\,{\sc i} 404.6\,nm and on the ratio between Ti\,{\sc ii}, Fe\,{\sc ii} 417.2-8\,nm, and Ti\,{\sc ii}, Fe\,{\sc ii} 444.4\,nm (see Fig.\,\ref{p2:fig:models}). 

\renewcommand{\arraystretch}{1.3}
\label{sec: stellar}
\begin{table}[]
\caption{Stellar parameters of PDS\,111.}
\centering
\begin{tabular}{l|ll|l}
\toprule \midrule
Star parameter    &    Value          & unit         & ref                \\ \midrule 
RA  & 05 24 37.25  & h\,m\,s & 1 \\
Dec & -08 42 01.71 & $^{\circ}\,^{\prime}\,^{\prime\prime}$ & 1 \\
$\mu_{RA}$ & $8.353\pm0.033$ & mas\,yr$^{-1}$ & 1 \\
$\mu_{Dec}$ & $-12.394\pm0.028$ & mas\,yr$^{-1}$ & 1 \\
Parallax & $6.3136\pm0.0352$ & mas & 1 \\
Distance & $157.85^{+0.95}_{-0.74}$ & pc & 4 \\
$v_{\text{rad}}$ & $25.8\pm0.1$ & km\,s$^{-1}$ & 3 \\
Spectral Type & G2\,V &  & 3 \\
$T_{\text{eff}}$ & $5900^{+100}_{-150}$ & K & 3 \\
log\,$g$ (spectrum)& $4.2\pm0.2$ & cm\,s$^{-1}$ & 3 \\
log\,$g$ (SED) & $4.13$ & cm\,s$^{-1}$ & 3 \\
$G$ & 9.693 & mag & 1 \\
$V$ & $9.97\pm0.22$ & mag & 2 \\
$B$-$V$ & $0.729\pm0.038$ & mag & 1 \\
Luminosity & $2.52^{+0.75}_{-0.40}$ & L$_{\odot}$ & 3 \\
Radius & $1.52^{+0.30}_{-0.17}$ & R$_{\odot}$ & 3 \\
E($B$-$V$) & 0.065  & mag   & 3 \\
$A_{\text{V}}$ & 0.2$^{+0.2}_{-0.1}$ & mag & 3 \\
$P_{\rm rot}$ & 3.1 & days & 3 \\
Evolutionary mass & $1.2^{+0.1}_{-0.1}$ & M$_{\odot}$ & 3 \\
Spectroscopic mass & $1.15-1.34$ & M$_{\odot}$ & 3 \\
Age & $15.9^{+1.7}_{-3.7}$ & Myr &  3\\
inclination & 58.2 & $^{\circ}$ & 3 \\
$v$\,sin$i$ & $21^{+0.4}_{-0.2}$ & km\,s$^{-1}$ & 3 \\

\midrule
\bottomrule
\end{tabular}
\tablebib{
1. \cite{Gaia2022} (ICRS, epoch 2000.0 calculated by Vizier),
2. \cite{Jayasinghe2018},
3. This work, 
4. \cite{BailerJones2021}.}
\label{p2:tab:star prop}
\end{table}
\renewcommand{\arraystretch}{1.}

The temperature, surface gravity, and rotational velocity were constrained through the fitting of Kurucz models \citep{kurucz1993} in the spectral range of 402-430\,nm with varying grid parameters based on the spectral classification; $T_\text{{eff}}$ (5500-6500\,K in steps of 100\,K), log\,$g$ (4.0-4.5 in steps of 0.1), and $v$\,sin$i$ (20-50\,km~s$^{-1}$ in steps of 1\,km~s$^{-1}$). The best fit model
is with a $T_\text{{eff}}$ of 5900\,K, log\,$g$ of 4.2, and $v$\,sin$i$ of 21\,km~s$^{-1}$; see Appendix\,\ref{p2:app:chi2} for the $\chi^2$ minimization results and Fig.\,\ref{p2:fig:models} for the best fit and sensitivity of the spectrum to changes in effective temperature. 
The $T_\text{{eff}}$ value is typical for G2-type pre-MS stars \citep{Pecaut2013} and the gravity is in agreement with luminosity class V.

Similar to \cite{woitke2016} and \cite{valegard2021}, the Kurucz-model spectrum was fitted to optical photometry to determine the star's luminosity, radius, extinction parameters 
, and log\,$g$ from the spectral energy distribution (SED) fit. The results are reported in Tab.\,\ref{p2:tab:star prop}. The optical photometry is known to vary \citep[see e.g.,][]{Torres1995, Rojas2008, zacharias2009, dasilva2009, kiraga2012}; therefore, we selected the mean V-band measurement of the ASAS-SN survey with 279 measurements over a period of five years in order to account for the variations \citep[$\Delta$V$\sim$0.43\,mag,][]{Jayasinghe2018}. A similar B-band study for PDS\,111 is lacking. We used Gaia magnitudes to inspect the behavior of $B$-$V$, applying Gaia photometry conversion tables from \cite{busso2022}. The mean $B$-$V$ is 0.729\,mag and varies with 0.075\,mag. Although these variations are significant compared to the average errors on the $B$-$V$ (<1\%), we neglected color variability for our use because the absolute variations of the V-band were substantially higher. Therefore, we used the mean $B$-$V$ from Gaia to obtain a B-band photometric point for the SED fit. For the fitting procedure, we fixed the distance to the Gaia distance from DR3\footnote{The Renormalized Unit Weight Error (RUWE) is 2.6 which is discussed in Sec.\,\ref{p2:sec:sed}.} \citep{Gaia2022} and effective temperature. The SED fit can be found in Appendix\,\ref{p2:app:sed}, which shows an infrared excess compared to the stellar SED. We might observe excess flux compared to the bluest photometric point from SDSS in the u-band ($\lambda_{\rm eff}$ = 360.804\,nm), but more photometry blueward from the u-band is needed to confirm this. 
We used the variability of the V-band magnitude in the ASAS-SN survey, extinction, effective temperature, distance, and bolometric correction from \cite{Pecaut2013} to infer errors on the luminosity and radius. 

The luminosity and $T_{\text{eff}}$ of the star result in a radius of 1.52\,R$_{\odot}$. Alternatively, the radius was derived from the $v$\,sin$i$, rotation period, and inclination of the star. The TESS photometry displays a dominant period with the use of a Lomb-Scargle periodogram \citep{lomb1976, scargle1982} of 3.137 days. This period is often attributed to the rotation period of the star. 
Using this period, the $v$\,sin$i$ of 21\,km\,s$^{-1}$ from Kurucz model fitting and an inclination of 58$^{\circ}$ (see Sec.\,\ref{sec:ALMA}), we calculated a radius of 1.53\,R$_{\odot}$, which is within errors similar to the derived radius from the SED fit.

\subsection{Scattered light observations} \label{p2:sec:scatter disk} \label{p2:sec:sphere results}
Our SPHERE H-band observations reveal an extended disk seen in polarized scattered light. We traced disk signal out to $\sim$1\arcsec\ from the central star along the major axis of the disk, corresponding to $\sim$158\,au. We show the SPHERE polarized and total intensity reductions in Fig.~\ref{fig: qphi-adi-main}. 
The polarized light image shows a classical flaring disk with an illuminated top side, a dark mid-plane and an illuminated outer edge of the disk bottom side (going from North to South in the image). The disk is seen at an intermediate inclination. Tracing of the outer edge of the top side of the disk by eye yielded an inclination of $\sim$60$^\circ$ at a position angle (PA) of the major axis of $\sim$68$^\circ$.

The total intensity image in Fig.~\ref{fig: qphi-adi-main} traces sub-structures on the disk top side. The employed angular differential imaging processing removed most of the signal with low spatial frequencies due to image self-subtraction (see e.g., \citealt{Milli2012, Ginski2016, Stapper2022}). We identified four pairs of arc-like structures, which indicate the presence of rings on the disk surface. As discussed in \cite{Ginski2016} the shape of these arcs is likely distorted due to the same image self-subtraction effects that filtered out the low spatial frequency data. In the rightmost panel of Fig.~\ref{fig: qphi-adi-main} we overlayed the contours of the total intensity image on the polarized light image which does not suffer from distortions of the disk morphology. The brightest pair of arcs in total intensity is tracing the outer edge of the disk top-side, while the two pairs of arcs further North trace rings in the inner part of the disk, also visible in the polarized light data. The faintest and southernmost pair of arcs in total intensity is tracing the edge of the disk mid-plane.

\begin{figure*}
\centering
    \tabcolsep=0.05cm 
    \begin{tabular}{cc}  
     \includegraphics[width=0.55\hsize]{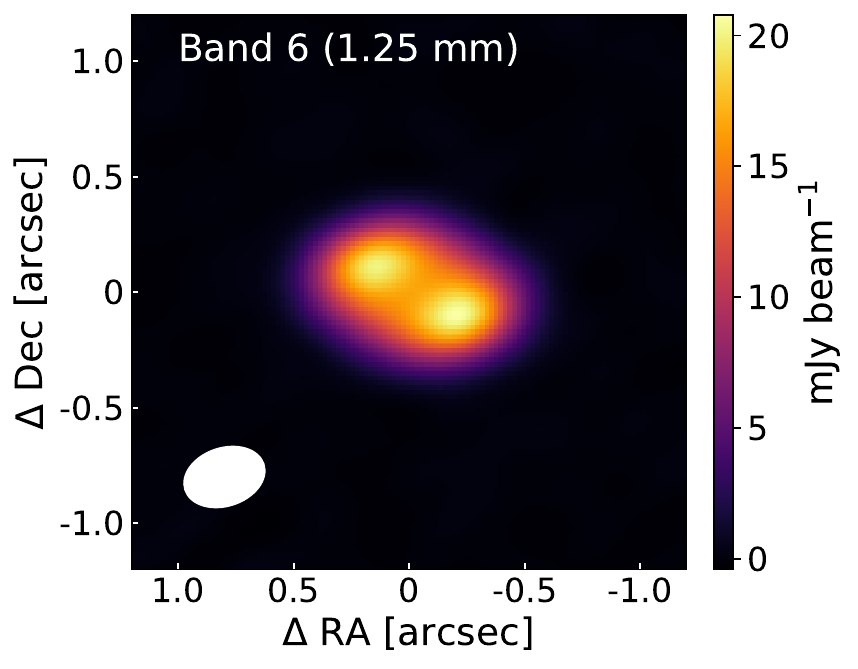}&
     \includegraphics[width=0.45\hsize]{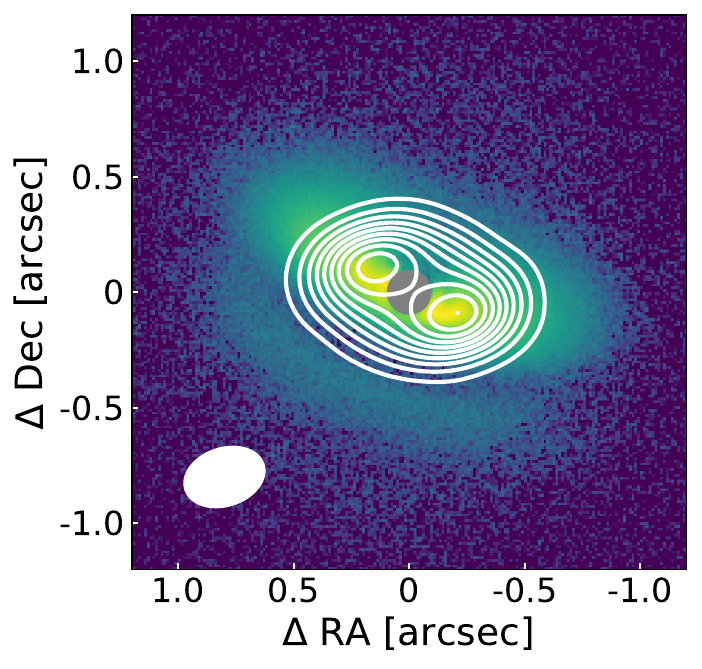}
    \end{tabular}
\caption{ALMA observations of PDS\,111. Left panel: Dust continuum emission at 1.25\,mm of PDS\,111. The resolution of the observations is 0.357''$\times$0.25'' with a position angle of -71.9$^{\circ}$ and is shown in the lower left part of the figure. Right panel: Overlap of ALMA (contours at 10,...100\% of the peak of emission) and SPHERE observations (in color). } \label{fig:ALMA_continuum}
\end{figure*}

\begin{figure*}
\centering
\includegraphics[width=16.0cm]{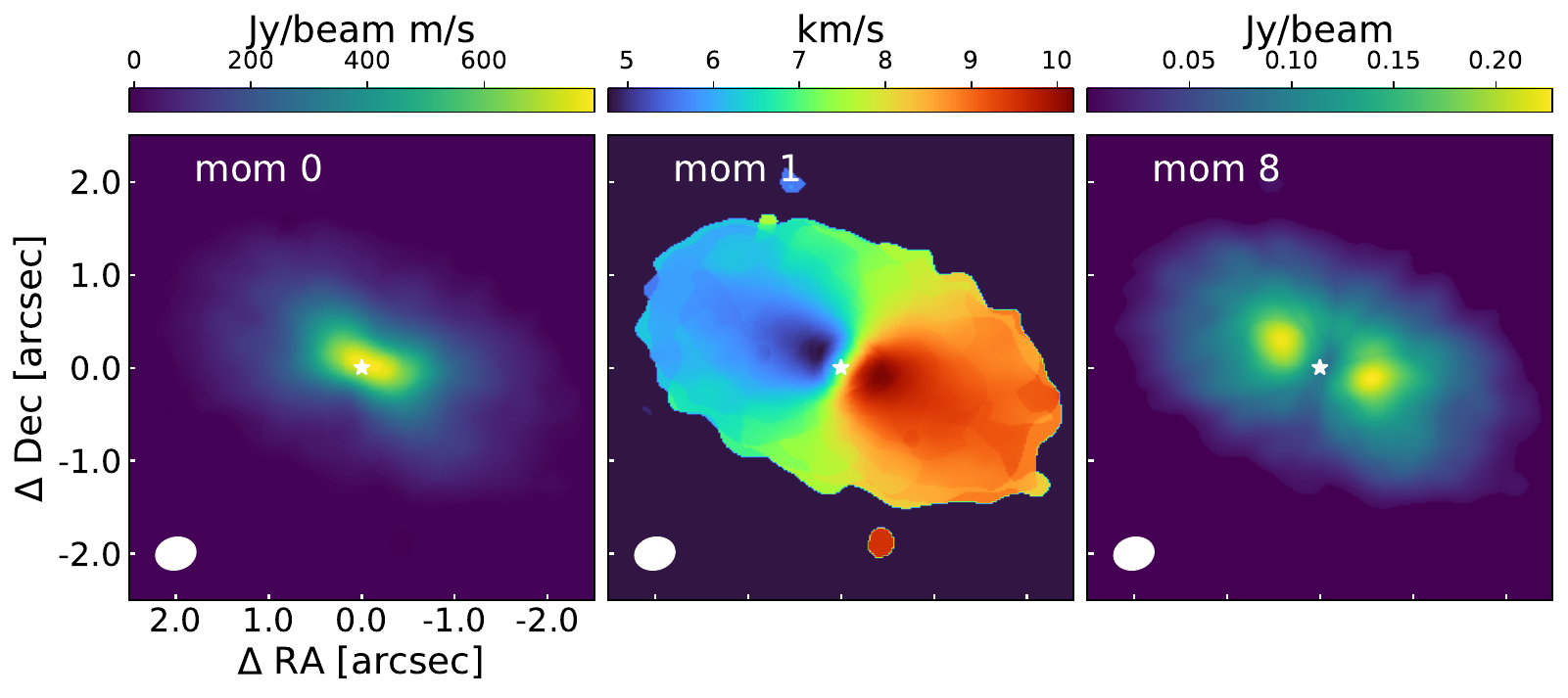}
\caption[]{Moment 0, 1, and 8 of the $^{12}$CO of PDS\,111. Moment 0 (left) is the integrated value of the spectrum, moment 1 (middle) the intensity weighted coordinate to get the velocity dispersion and moment 8 (right) shows the peak value of the spectrum. The beam size is 0.429\arcsec\ $\times$ 0.342\arcsec\ with a PA of -76$^{\circ}$.114, shown as the oval in the lower left corner of the panels.}
\label{fig:12CO_moments}
\end{figure*}

\subsection{Dust and line observations}\label{sec:ALMA}

\subsubsection{Dust continuum emission}
Fig.~\ref{fig:ALMA_continuum} shows the dust continuum emission observed with ALMA with a final beam of 0.357''$\times$0.25'' with a PA of -71.9$^{\circ}$ at 1.25\,mm after self-calibration as described in Sec.~\ref{ALMA_obs}. To check if the emission is optically thin, we calculated the optical depth of the peak of the continuum by assuming \citep[e.g.,][]{dullemond2018}:

\begin{equation} \label{eq:opticaldepth}
\begin{aligned}
    & \,I_\nu(r)=B_\nu(T_\mathrm{d}(r))(1-\exp{[-\tau_\nu(r)]}) \quad \mathrm{thus} \\
    &\, \tau = - \ln \left(1-\frac{I (r_\mathrm{peak})}{B(T_d(r_\mathrm{peak}))} \right),
\end{aligned}
\end{equation}

\noindent with $T_d$ being the dust temperature at the peak location ($r_{\rm{peak}}$). Assuming a dust temperature of 20\,K, the optical depth at the peak is $\tau_{\rm{peak}}=0.27$, similar to the values found in the DSHARP sample \citep[][]{huang2018, dullemond2018}. This value of the optical depth may be underestimated because dust scattering could be neglected \citep{zhu2019}, which would result from unresolved substructures in the ring, and/or a lower dust temperature or the large beam size relative to the disk.

If the emission is optically thin, the dust disk mass can be calculated as 
\begin{equation}
    M_{\mathrm{dust}}\simeq\frac{{d^2 F_\nu}}{\kappa_\nu B_\nu (T)},
\end{equation}

\noindent where $d$ is the distance to the source, $F_\nu$ is the total flux at 1.3\,mm, and $B_\nu$ is the black-body surface brightness at a given temperature \citep{hildebrand1983}. Taking for the mass absorption coefficient $\kappa_\nu=2.3\,$cm$^{2}$\,g$^{-1}\times(\nu/230\,\rm{GHz})^{0.4}$ \citep{beckwith1990, andrews2013}, we calculated a dust disk mass of 45.8\,M$_\oplus$ or $\sim0.14$\,M$_{\rm{Jup}}$, similar to typical disks around T\,Tauri stars in different star-forming regions \citep[e.g.,][]{pascucci2016}.

\subsubsection{$^{12}$CO emission}

Fig.~\ref{fig:channel_maps} presents the channel maps of $^{12}$CO of PDS\,111 observed with ALMA, showing emission from $\sim$0.4\,km~s$^{-1}$ to $\sim$14.7\,km~s$^{-1}$. For comparison, the $5\sigma$ level of the continuum emission is overplotted, and it shows that $^{12}$CO is more extended than the dust continuum emission and the scattered light detection.

Fig.~\ref{fig:12CO_moments} shows the moment 0, 1, and 8 (peak value of the spectrum) of the $^{12}$CO of PDS\,111 generated with a Keplerian mask and clipping at $3\sigma$ of the JvM corrected image as described in Sec.~\ref{ALMA_obs} (the beam size is 0.429\arcsec\ or 0.342\arcsec\ with a PA of -76$^{\circ}$.114). To extract the stellar mass from the $^{12}$CO Keplerian rotation pattern, we used \texttt{eddy}\footnote{Tool to extract kinematic information from spatially and spectrally resolved line data}, \url{https://eddy.readthedocs.io/en/latest/} \citep{eddy}, which gives 1.29$^{+0.82}_{-0.18}$\,M$_{\odot}$ when assuming a simple disk model and fixing the inclination to be that of the continuum. The large uncertainty in the stellar mass may come from the unknown vertical structure and disk orientation. Higher angular resolution is needed to better resolve the surface height with ALMA.

\section{Age and disk properties}\label{p2:sec:age_and_disk}
\begin{figure*}
\centering
\includegraphics[width=\hsize]{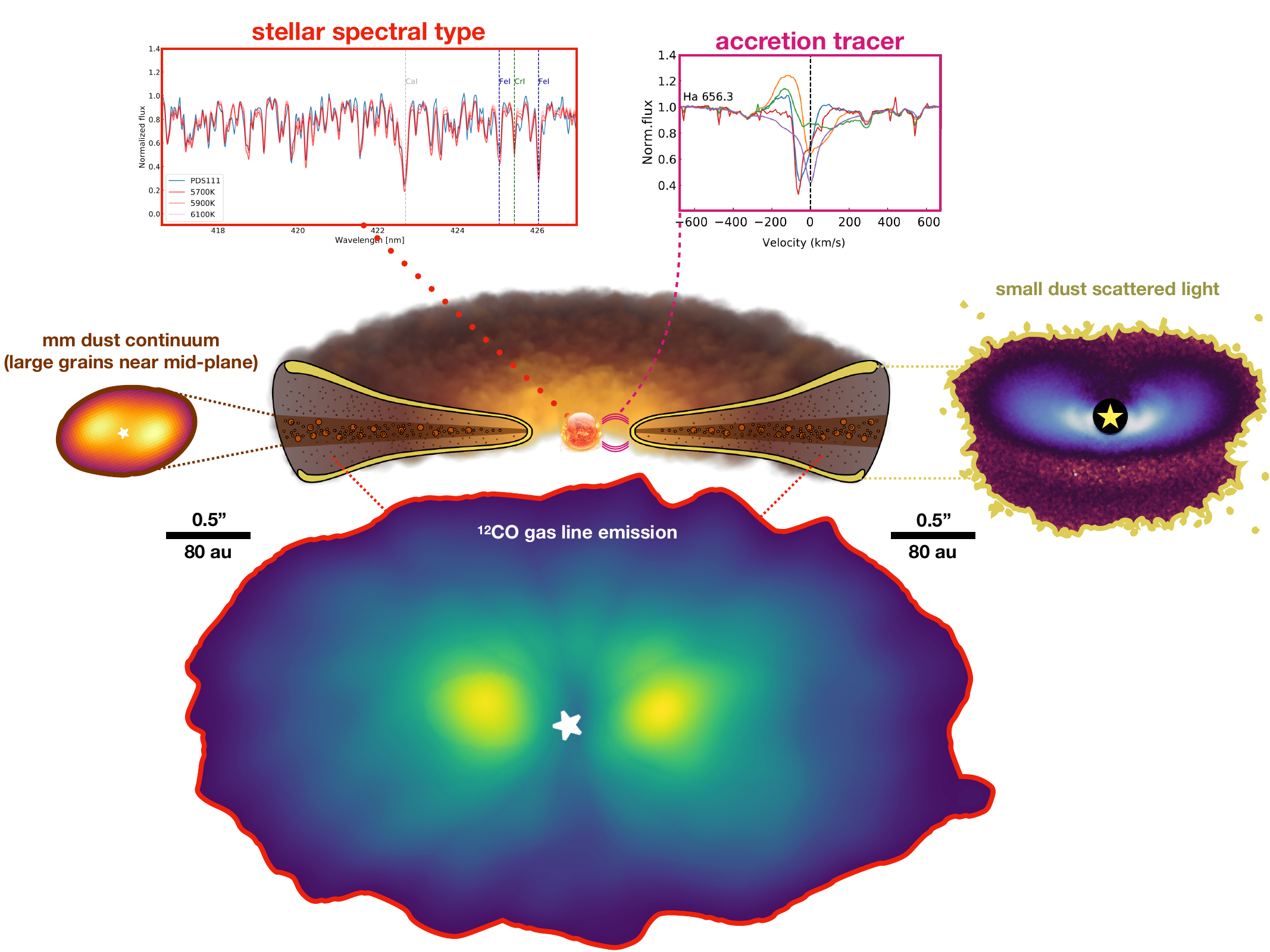}
\caption[]{Overview of PDS\,111 indicating the parts of the system that are traced by the different observational techniques.
}
\label{p2:fig:disk_overview}
\end{figure*}

By combining the observations at different wavelengths, we are able to sketch a more complete picture of the system. Fig.\,\ref{p2:fig:disk_overview} shows what different parts of the disk are traced by the different observing techniques. 

\subsection{Age}\label{p2:sec:sed}
\begin{figure}
\centering
\includegraphics[width=\hsize]{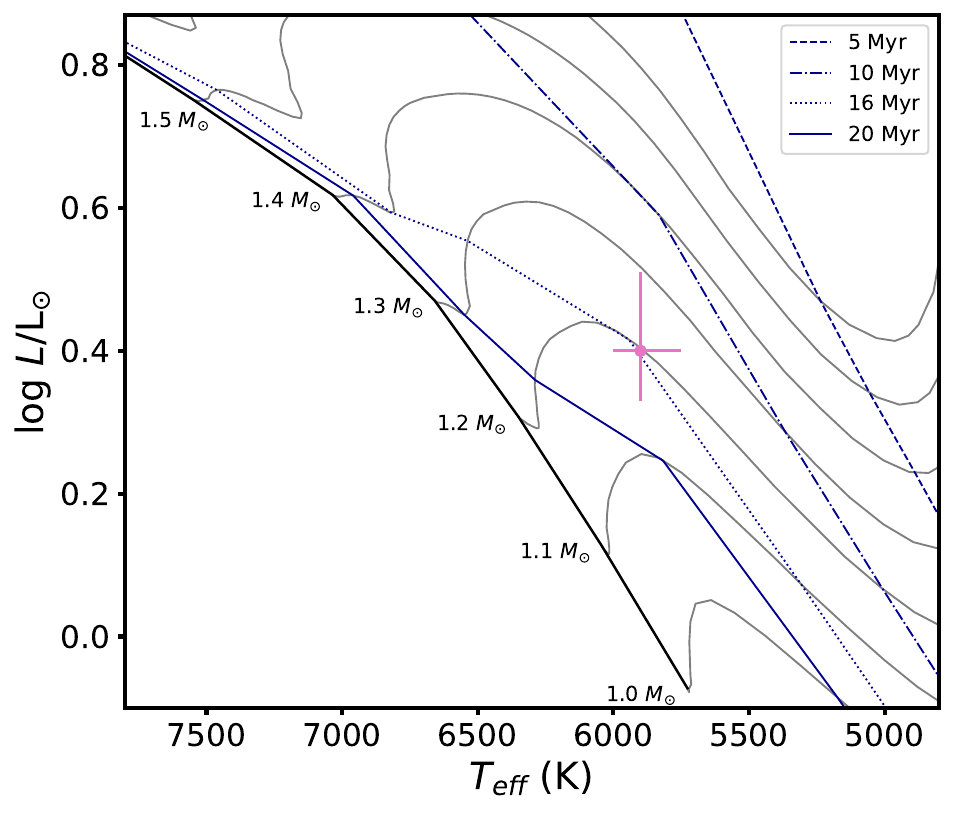}
\caption[]{Hertzsprung-Russell diagram with PMS tracks and isochrones from \cite{siess2000}. The location of PDS\,111 coincides with an age of 15.9$^{+1.7}_{-3.7}$\,Myr.}
\label{p2:fig:hrd}
\end{figure}

Using the luminosity and effective temperature, we placed the star in the Hertzsprung Russell Diagram (HRD) on evolutionary tracks from \cite{siess2000}, and infer an age of 15.9\,Myr and a evolutionary mass of 1.20\,M$_{\odot}$, see Fig.\,\ref{p2:fig:hrd}.
The error on the effective temperature and conservative error on the luminosity lead to an uncertainty range of $12.2-17.6$\,Myr for the age and $1.15-1.30$\,M$_{\odot}$ for the mass. The mass is similar to the best fit mass value obtained in Sec.~\ref{sec:ALMA}. Additionally, using the radius and surface gravity of the star in Tab.\,\ref{p2:tab:star prop}, we measured the spectroscopic mass. Using the log\,$g$ from the SED fit resulted in a mass of 1.14\,M$_{\odot}$ and using the log\,$g$ from the spectrum fit resulted in a mass of 1.34\,M$_{\odot}$. Both are within the errors consistent with the evolutionary mass.

\cite{manara2013} show that the age of T\,Tauri stars could be over-estimated when the extinction, accretion, and PMS stellar models are not fitted simultaneously. However, their method cannot be applied to PDS\,111 since their stellar models go up to spectral type G4. The accretion rate is measured from line luminosities in Sec.\,\ref{sec:hot disk} demonstrating that the accretion rate for PDS\,111 is low for a star of this mass \citep{2023ASPC..534..539M}.

The estimated age of the system of 15.9\,Myr could be wrong if the distance to PDS\,111 obtained from the {\it Gaia} DR3 parallax \citep{Gaia2022} is not correct. This may be the case considering the large RUWE value of 2.6, despite the small error on the parallax (0.035~marcsec). The RUWE statistic shows whether a single-star light profile fits the observations. The circumstellar material of PDS\,111 might inflate its value. \cite{fitton2022} determine that for disk-bearing sources a RUWE cutoff value of 2.5 is more appropriate. The RUWE value of PDS\,111 is slightly above this cutoff\footnote{The circumstellar disk may inflate the RUWE value.}. We checked the {\it Gaia}-database to see whether the \texttt{visibility\_period\_used} is more than 10 and \texttt{ipd\_gof\_harmonic\_amplitude} is less than 0.1. Both is the case: the former is 22 and the latter is 0.029. Moreover, the {\it Gaia} parallax cross-matches that obtained with {\it Hipparcos} \citep{Gaia2022, hipparcos1997}. Therefore, we do not find any obvious indications that the {\it Gaia} measurements are inaccurate.

Throughout the SED fitting procedure, the total-to-selective visual extinction $R_{\rm V}$ is fixed to 3.1 \citep{Cardelli1989}. However, it has been noticed that $R_{\rm V}$ is sometimes higher toward star-forming regions \citep[e.g.,][]{Ramirez-Tannus2017}. If this value would for example be 4.1, the extinction toward the source would be higher and the luminosity would increase to a value of 3.6\,L$_{\odot}$ and the corresponding age in the HRD would be 11.3\,Myr for the same temperature, lowering the age of the system. More simultaneous photometric measurements over a period of time could help to improve the SED fit and fit for a range of different $R_{\rm V}$ values in order to find the best fitting parameter in the sightline of PDS\,111. We conclude that the uncertainty in distance or extinction does not reject our hypothesis that PDS\,111 is relatively old, that is at least $\sim$11\,Myr. This relatively old age is similar to what has been determined for the stars that are surrounded by a "Peter Pan Disk"; we further discuss this scenario in Sec.\,\ref{p2:sec:peterpan}.

\begin{figure}
\centering
\includegraphics[width=0.99\hsize]{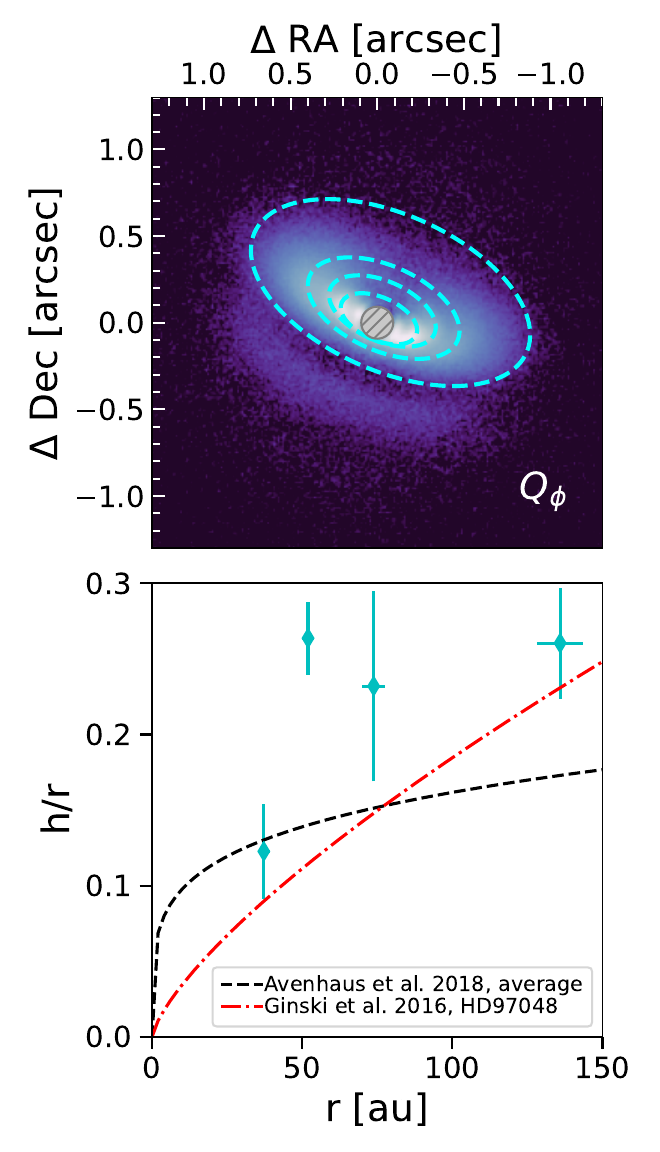}
\caption{Ellipses fitted to the radial sub-structures traced in scattered light (top) and resulting height of the scattering surface of the disk (bottom). We compare the scattering surface height with the average surface height profile of several T\,Tauri stars found by \cite{Avenhaus2018} and the extreme flaring case of the Herbig star HD\,97048, discussed by \cite{Ginski2016}.}
\label{fig: scale-height}
\end{figure}

\begin{figure}
\centering
\includegraphics[width=0.99\hsize]{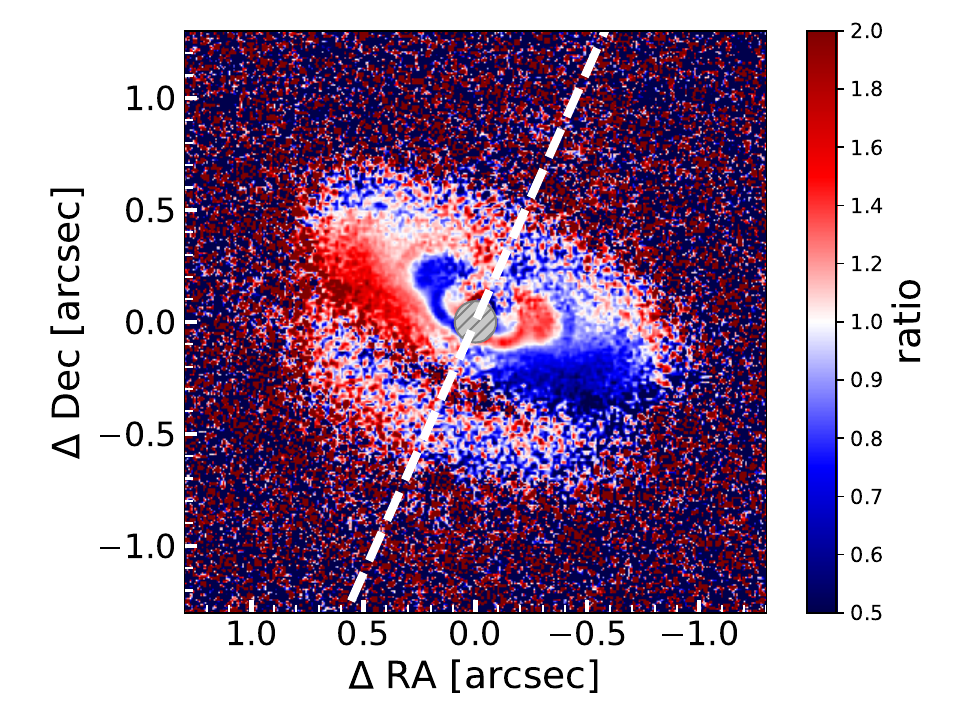}
\caption[]{The scattered light disk flipped around the minor axis and divided. The minor axis is indicated by the white, dashed line. The scattered light disk is asymmetric relative to the minor axis. The colorbar gives the flux ratio between the two sides of the disk, showing that the flux deviates up to a factor of two between both sides. Notably, the flux ratio turns around for the region inside of $\sim$0.3\arcsec\ along the major axis. This may indicate a local change in scale height symmetric around the disk midplane rather than a warp or misalignment since the bottom side displays a similar flux ratio.
}
\label{fig: sphere-asymmetry}
\end{figure}

\subsection{Disk surface}

As discussed in \cite{deBoer2016}, \cite{Ginski2016}, and \cite{Avenhaus2018}, we used the identified disk sub-structures to extract the height profile of the disk surface layer at which the disk becomes optically thick for NIR radiation. To facilitate the tracing of the structures we employed high pass filtering as well as deconvolution techniques using the Adaptive Image Deconvolution Algorithm (AIDA) code by \cite{Hom2007}. We show the high-pass filtered and the de-convolved data in Fig.~\ref{fig: app: ellipse} in Appendix~\ref{app:ellipse}. Because the disk structure is complex, we traced the individual rings by eye as was done by \cite{Avenhaus2018} for the case of IM\,Lup. Tracing was done for all structures in the original polarized light images, in the high-pass filtered images and in the de-convolved images to ensure that there was no mismatch in structures. To reduce the number of free parameters, the inclination of the individual rings as well as their PA were fixed to the values extracted from the ALMA data discussed in Sec.~\ref{sec:ALMA}. Thus, for each ring we traced an ellipse by re-scaling to match the semi-major axis and then applying an offset along the disk minor axis to match the height profile of the disk. Our results are shown in Fig.~\ref{fig: scale-height} (top). Additionally we show a saturated image of the disk for the overlay of the ellipse on the outer disk edge in Fig.~\ref{fig: app: ellipse-saturated}.

We overlay the resulting ellipses on the polarized light image and calculate the disk aspect ratio from the ellipse center-offset along the minor axis and the ellipse semi-major axis following \cite{deBoer2016}. We find a steep increase in the disk aspect ratio from the innermost ring at 37\,au to the second ring at 52\,au from 0.12 to 0.26. While the aspect ratio of the inner ring is compatible with results presented by \cite{Avenhaus2018} for a small sample of T\,Tauri stars, the subsequent structures all show a higher aspect ratio, indicative of a strong flaring geometry of the disk, see Fig.\,\ref{fig: scale-height} (bottom). In addition to the two rings at 37\,au and 52\,au we identify a third more diffuse ring-structure in the deconvolved and high-pass filtered images (see figure~\ref{fig: app: ellipse}) at a separation of 74\,au. Since this structure is of lower spatial frequency than the inner two rings it is entirely filtered out by the ADI post-processing and thus not visible in the total intensity image in Fig.~\ref{fig: qphi-adi-main}. We tentatively overlay an ellipse on this ring as well and find an aspect ratio of 0.23, roughly consistent with the inner ring at 52\,au, however, with larger uncertainties.

The increase of the aspect ratio between the innermost ring at 38\,au and the outer disk edge at 140\,au, for which we found a value of 0.26, is in principle compatible with the extreme flaring profile found for the disk around the Herbig star HD\,97048 by \cite{Ginski2016}. However, since the aspect ratio at intermediate separations from the central star appears to be significantly higher, we could not fit the entire surface height profile with a single power law. As briefly discussed in \cite{Ginski2016}, this may be due to an optical depth effect, disproportionately affecting the outer disk region. As the dust density in the outer disk decreases, the optically thick layer sinks closer to the mid-plane.

\subsection{Disk asymmetry}\label{sec:disk asymmetry}

For a disk with an azimuthally homogeneous dust and gas distribution, we would expect that the polarized light image is symmetric with respect to the disk's minor axis. However, this does not appear to be the case for PDS\,111. To highlight the disk asymmetry we flipped the image around the minor axis and then divided the original image by the flipped image. This highlights the areas in which both sides of the disk deviate. We show the result in Fig.~\ref{fig: sphere-asymmetry}. We see two main effects; first of all the disk is indeed asymmetric with flux deviations of up to a factor two between both sides. Second, we find a flip in the asymmetric brightness distribution (the white line between the red and blue areas) around 0.3\arcsec\ (47\,au) along the major axis. Inside of this border the North-East (NE) side of the disk is fainter than the South-West (SW) side, while it is the opposite way in the outer disk.

Since the SPHERE observations are tracing scattered stellar light, the disk brightness is intricately linked to the disk height profile and geometry. Broad brightness asymmetries as observed here can be caused by warps or misalignment in inner disk regions and subsequent shadowing of the outer disk (e.g., \citealt{Debes2017,Benisty2018,Muro-Arena2020}). In such a scenario the breaking or warping point would be at the location where we observe the flip in the asymmetric signal of the disk. In that case, the scale height inside of the flipping point would be increased in the SW on the disk top and decreased in the NE. This could then explain simultaneously why the disk is brighter inside the flipping point (it is more exposed to stellar light) and fainter outside the flipping point (the outer disk is shadowed by the inner disk) in the SW.

However, we note that in a warping scenario we would expect that the disk bottom side shows the opposite behavior from the disk top side, that is, we would expect the disk bottom side to be bright in the SW. While the S/N is lower on the disk bottom, it nevertheless appears that it instead exhibits the identical behavior as the disk top side, which means it is faint in the SW and bright in the NE. This may then indicate that the disk is not radially warped or misaligned as a whole but rather shows a local change in scale height, symmetric around the disk mid-plane. This may be compatible with the fact that we are measuring an increased aspect ratio for the disk ring at 52\,au. However, since we traced the ring at all azimuthal angles simultaneously a localized effect may be averaged out.

\subsection{Disk structure}\label{sect:visi}

The dust continuum emission morphology was analyzed with a parametric brightness modeling using the packages \texttt{galario} \citep{galario} to calculate the Fourier Transform of the models and \texttt{zeus} \citep{karamanis2020, karamanis2021} to run an MCMC and sample the posterior likelihood of the parameter space. We considered four components to describe the PDS\,111 emission: A centrally peaked Gaussian ($g_1$) motivated by the inner emission found in several single-ringed disks \citep[e.g.,][]{pinilla2021, ackerman2021}, two Gaussian profiles for the ring ($g_2$, $g_3$) to allow for radially asymmetric ring morphology (rather than one broken Gaussian profile), and a Gaussian azimuthal asymmetry ($g_4$) to describe the over-emission observed in the South-western side of the disk. Fig.~\ref{fig:components} illustrates each of the components of the model used to fit the morphology of PDS\,111. We also include as free parameters the central position of the disk ($x_0$, $y_0$) and the geometry, given by inclination and $\rm PA$. Every parameter has a flat prior and boundaries representative of physical limitations, such as only allowing positive distances and brightness. The model images have a pixel size of 4\,mas and an image size of 2048 pixels. We ran the MCMC with eight times the number of parameters as the number of walkers ($n_{\text{walk}}=136$). For each walker, the burn stage was 5000, and 6000 for the parameter space sampling. The summary of the modeling results is given in Tab.~\ref{tab:disk_prop}.

\begin{figure*}
    \centering
    \includegraphics[width=16.0cm]{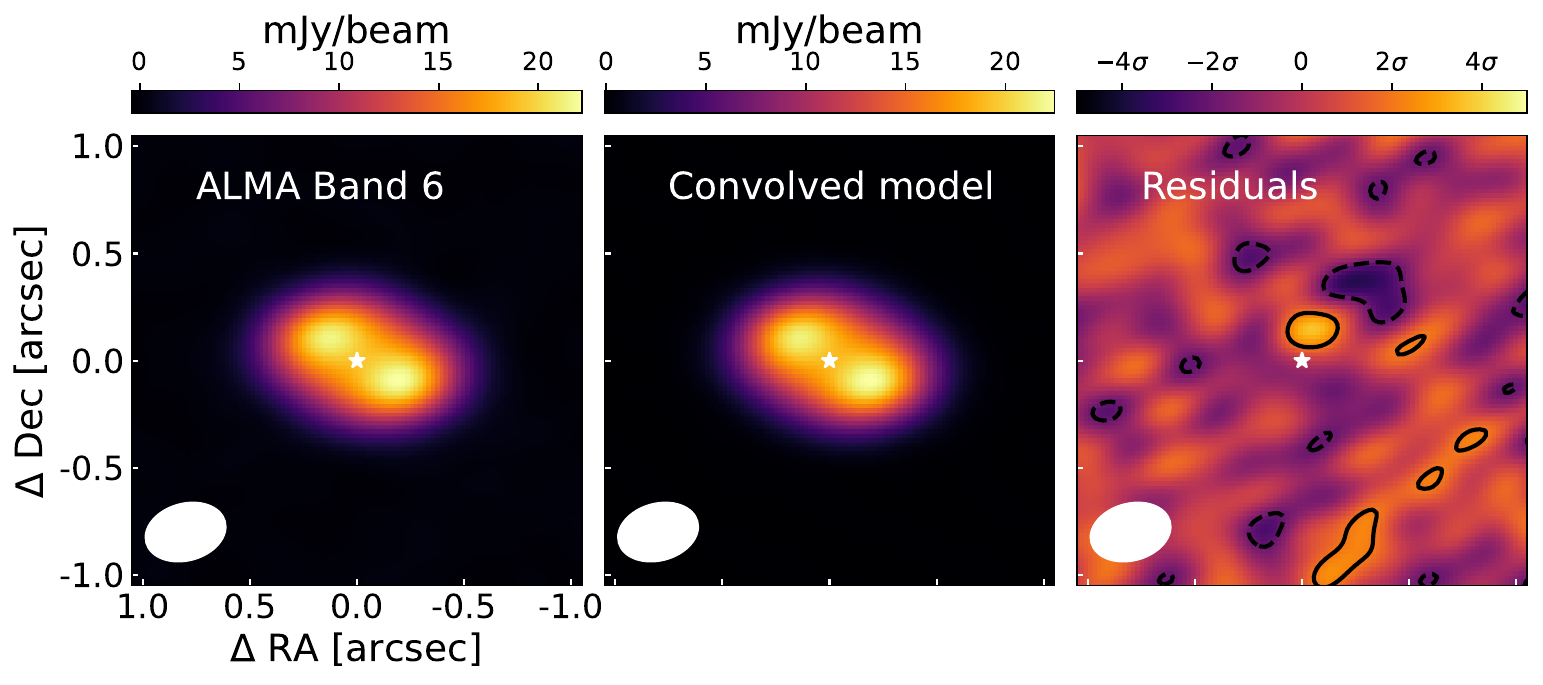}
    \caption{Observations versus model images from our best fit with \texttt{galario} after convolving with the same beam as the observations and the residuals. The contours of the residual map are [$-4\sigma$, $-2\sigma$, $2\sigma$, $4\sigma$], where negative values are dashed contours and positive values solid contours.}
    \label{fig:galario_fit}
\end{figure*}

Fig.~\ref{fig:galario_fit} shows the comparison between the dust continuum observations and model images from our best fit with \texttt{galario} after convolving with the same beam as the observations and the residuals. The residual map demonstrates that our model recovers most of the morphology of the dust continuum emission. The main characteristic of the dust morphology is a cavity with a ring-like emission peaking at $\sim$30\,au (see Fig.~\ref{fig:components} for the model components). The radius that encloses 90\% of the total flux from the visibility fitting is $\sim$55\,au.

\renewcommand{\arraystretch}{1.3}
\begin{table}[t]
\caption{Dust continuum visibility modeling results.  }
\centering
\begin{tabular}{ c|c|c|c } 
  \hline
  \hline
\noalign{\smallskip}
           & Property      & Best value $\pm 1\sigma$ & unit \\
\noalign{\smallskip}
  \hline
\noalign{\smallskip}
Disks      & $x_0$ & $27.3 \pm 0.6$ & mas \\
Geometry   & $y_0$ & $ 7.4 \pm 0.3$ & mas \\
           & $inc$ & $58.2 \pm 0.1$ & deg \\
           & $PA$  & $66.2 \pm 0.1$ & deg \\
\noalign{\smallskip}
  \hline
\noalign{\smallskip}
$g_1$      & $f_1$      & $119.8_{-94.4}^{+146.5}$  & $\mu$Jy/pix \\
           & $\sigma_1$ & $  3.8_{- 2.6}^{  +4.9}$  & mas \\
\hline
$g_2$      & $f_2$      & $  1.8_{- 0.9}^{+ 1.3}$  & $\mu$Jy/pix \\
           & $r_2$      & $280.8_{-28.0}^{+45.0}$  & mas \\
           & $\sigma_2$ & $ 94.9_{-21.9}^{+25.3}$  & mas \\
\hline
$g_3$      & $f_3$      & $ 22.1_{- 0.7}^{+ 1.4}$  & $\mu$Jy/pix \\
           & $r_3$      & $286.6_{- 3.5}^{+ 2.6}$  & mas \\
           & $\sigma_3$ & $ 33.8_{- 6.1}^{+ 4.7}$  & mas \\
\hline
$g_4$      & $f_4$              & $  2.4_{- 0.7}^{+ 1.4}$  & $\mu$Jy/pix \\
           & $r_4$              & $286.7_{-54.2}^{+31.0}$  & mas \\
           & $\sigma_{r4}$      & $116.5_{-30.8}^{+ 4.7}$  & mas \\
           & $\theta_4$         & $-175.2 \pm 2.4$        & deg \\
           & $\sigma_{\theta4}$ & $ 11.3_{- 4.0}^{+ 3.6}$  & deg \\
\noalign{\smallskip}
  \hline
\noalign{\smallskip}
Properties & $F_{\text{tot}}$ & $139.1_{-0.6}^{+0.2}$  & mJy \\
           & $R_{68\%}$       & $307.2_{-1.7}^{+2.5}$  & mas \\
           & $R_{90\%}$       & $345.4_{-5.8}^{+0.9}$  & mas \\
           & $R_{95\%}$       & $376.6_{-11.1}^{+5.9}$ & mas \\
\noalign{\smallskip}
  \hline
  \hline
\end{tabular}
\label{tab:disk_prop}
\tablefoot{Uncertainties represent the $68\%$ dispersion range. Pixel size was 4\,mas. The central angle of the asymmetry ($\theta_4$) is measured relative to the disk $PA$. The second column gives the properties of the four Gaussian model components; flux (f), sigma ($\sigma$), radius (r), and asymmetry angle ($\theta$).}
\end{table}

\subsection{Radial extent of the disk}\label{p2:sec:radialprof}
\begin{figure}
\centering
\includegraphics[width=0.95\hsize]{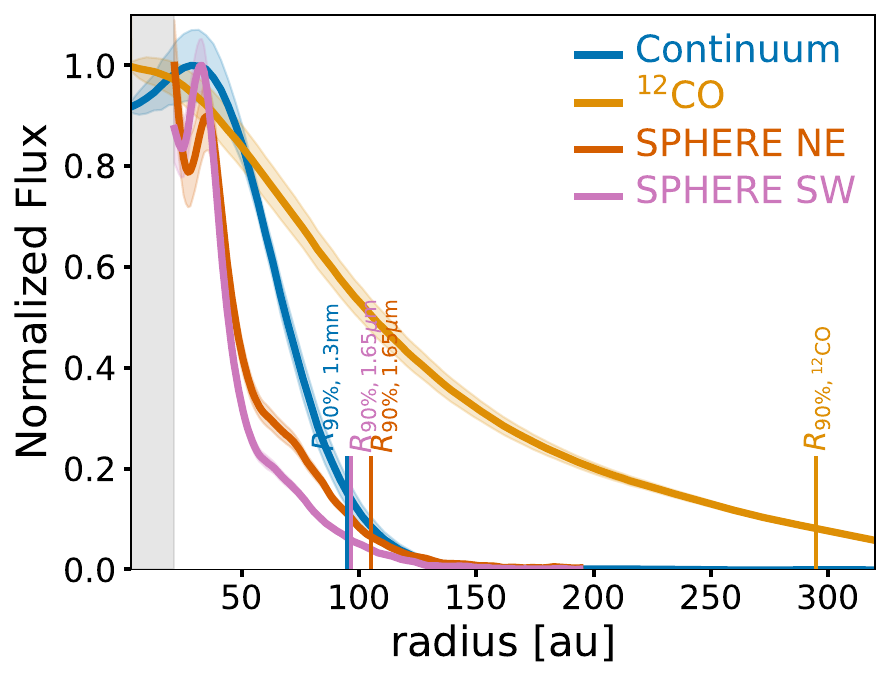}
\caption[]{Azimuthally averaged radial intensity profiles of the deprojected images from ALMA dust continuum and the moment 0 maps of the $^{12}$CO from ALMA Band-6 observations of PDS\,111. Each profile is normalized to the peak. The ALMA profiles include the standard deviation of each elliptical bin divided by the square root of the number of beams spanning the full azimuthal angle at each radial bin. For the SPHERE observations, we take two radial cuts at the North-East (NE) and South-West (SW) of the major axis of the disk. The vertical lines show the radius that encloses 90\% of the emission in each of the maps. The gray region shows the size of the coronagraph in the SPHERE observations.}
\label{fig:radial_profile}
\end{figure}

The radial distribution and extension of PDS\,111 in dust continuum, scattered light and $^{12}$CO is shown in Fig.~\ref{fig:radial_profile}. For the ALMA profiles, the images have been deprojected and then azimuthally averaged, while for the SPHERE data we show radial cuts at the North-East (NE) and South-West (SW) of the major axis of the disk. The vertical lines show the radius that encloses 90\% of the emission in each of the maps. The 90\% radius of the $^{12}$CO map ($\sim$295\,au) is much more extended than that tracing the dust, both in scattered light and continuum emission. The 90\% radius of the dust continuum emission is at 95\,au, three times smaller than observed in $^{12}$CO (both profiles being affected by convolution of a similar beam). The latter is very similar to the average ratio observed in several protoplanetary disks that are on average younger \citep{long2022}.

The radius that enclose 90\% of the flux measured directly from the dust continuum emission (95\,au) and that measured from the visibility fitting of the dust continuum emission (55\,au, Sec.\,\ref{sect:visi}) are different. This is a natural result of the large ALMA beam and the high inclination of the source. The $\sim$55\,au estimate is probably more representative of the actual size, therefore of the distribution of  mm-grains that dominate the emission at ALMA wavelengths. Fig.\,\ref{fig:radial_profile} shows that this mm-grain thermal emission disk is smaller than the disk size from scattered light ($\sim$105\,au for the NE and $\sim$96\,au for the SW cuts), which would indicate that the small grains responsible for the scattering are more radially extended than the large grains, as a product of radial drift \citep[e.g.,][]{guilloteau2011, tazzari2016, trapman2019, trapman2020}.
 
When the radial profiles from scattered light are multiplied by $r^{2}$ to compensate for the stellar illumination, the 90\% radius shifts to $\sim$155\,au for the NE and $\sim$136\,au for the SW emission (in line with the 158\,au mentioned in Sec.~\ref{p2:sec:sphere results} at which the disk signal drops below the noise floor along the major axis). The size of the disk in scattered light is thus $\sim$2.4-2.7 times larger than in the millimeter emission. Such a difference is similar to that found in the ALMA images of dust vs. gas emission. \cite{trapman2019, trapman2020} propose that optical depth effects cannot explain such a big difference, and this is likely evidence for small and large grain separation through radial drift. 

\subsection{Inner gaseous disk}\label{sec:hot disk}
The presence of hot gaseous circumstellar material is revealed by H$\alpha$ and H$\beta$ emission and strong variability in these and other H\,{\sc i}, He\,{\sc i}, Ca\,{\sc ii} and Na\,{\sc i} lines. We do not detect any forbidden emission lines. The equivalent width (EW) is determined after subtracting the best fit Kurucz model from the observations. This ensures that the measured EW is an indication of the emission or absorption from the circumstellar material. The EW measurements are listed in Tab.\,\ref{tab:EW}. The EW(H$\alpha$) emission is $<10$\AA, which classifies PDS~111 as a weak-line T~Tauri star. 

Following \citet{alcala2017}, we determined line luminosities by measuring the EW of the lines, the absolute flux from the SED in H$\alpha$ and H$\beta$ and converting this to luminosity using the distance to the star. Their empirical relations were used to convert the line luminosities to accretion luminosities and the resulting mass-accretion rates assuming a typical inner radius of 5\,R$_{\rm \star}$, are reported in Tab.\,\ref{tab:Macc}. This yields a rate that varies between 1-5$\times$10$^{-10}$\,M$_{\odot}$\,yr$^{-1}$. Due to the lack of photometric observations in the UV, the mass-accretion rate could be constrained only indirectly.

\begin{table}[]
\centering
\caption{Derived mass-accretion rates of H$\alpha$ and H$\beta$ at each epoch in log\,$\dot{\rm M}_{\rm acc}$ [M$_{\odot}$\,yr$^{-1}$].}
\begin{tabular}{l|llll}
\toprule \midrule
                $\lambda_{\rm rest}$ (nm)  & 28-11-'21 & 07-12-'21 & 20-10-'22 & 28-10-'22 \\
                 \midrule
H$\alpha$    656.28   & -9.47 & -9.28   & -9.37   & -9.69    \\
H$\beta$     486.14   & -9.42   &       & -10.02 & -9.42  \\
\midrule  
\bottomrule
\end{tabular}
\label{tab:Macc}
\end{table}

The temporal variance method \citep[TVS;][]{Fullerton1996,dejong2001} is employed to study spectroscopic variability. The observed standard deviation $\sigma_{\rm obs}$ is compared to the expected standard deviation $\sigma_{\rm exp}$ at a certain wavelength bin $j$. If the ratio of the two standard deviations exceeds unity, the variation in the spectral line between the epochs is larger than what is expected from the noise in that wavelength bin. Likewise, when no variation is detected, the ratio (i.e., the value of the TVS) is randomly distributed around unity, reflecting the noise in the data.

Similar to \cite{dejong2001} and \cite{2024A&A...681A.112D}, $\sigma_{{\rm obs},j}$ is determined from the average of the standard deviations of the bins in the continuum region around the spectral line in all epochs. This yields $\sigma_{\rm con}$, which is scaled by the ratio of the mean normalized flux of all epochs of the evaluated wavelength bin $j$ and the continuum. The $\sigma_{{\rm obs},j}$ is calculated from observed fluxes in wavelength bin j. The TVS value is given by the ratio: 
\begin{eqnarray}
{\rm TVS}_{j} = \frac{\sigma_{{\rm obs}, j}}{\sigma_{{\rm exp}, j}} = \frac{\sigma_j}{\sigma_{\rm con}} \sqrt{\frac{\bar{F}_{\rm con}}{\bar{F}_j}}.
\end{eqnarray} Here $\bar{F}_j$ is the average flux in wavelength bin $j$ for all epochs and $\bar{F}_{\rm con}$ the average continuum flux. The significance of the variability is determined by comparing a $\chi^2$-distribution with the degrees of freedom set to the number of observations, to the TVS value. A 95\% level is used for a significant detection.

\begin{figure*}
\centering
\includegraphics[width=0.9\hsize]{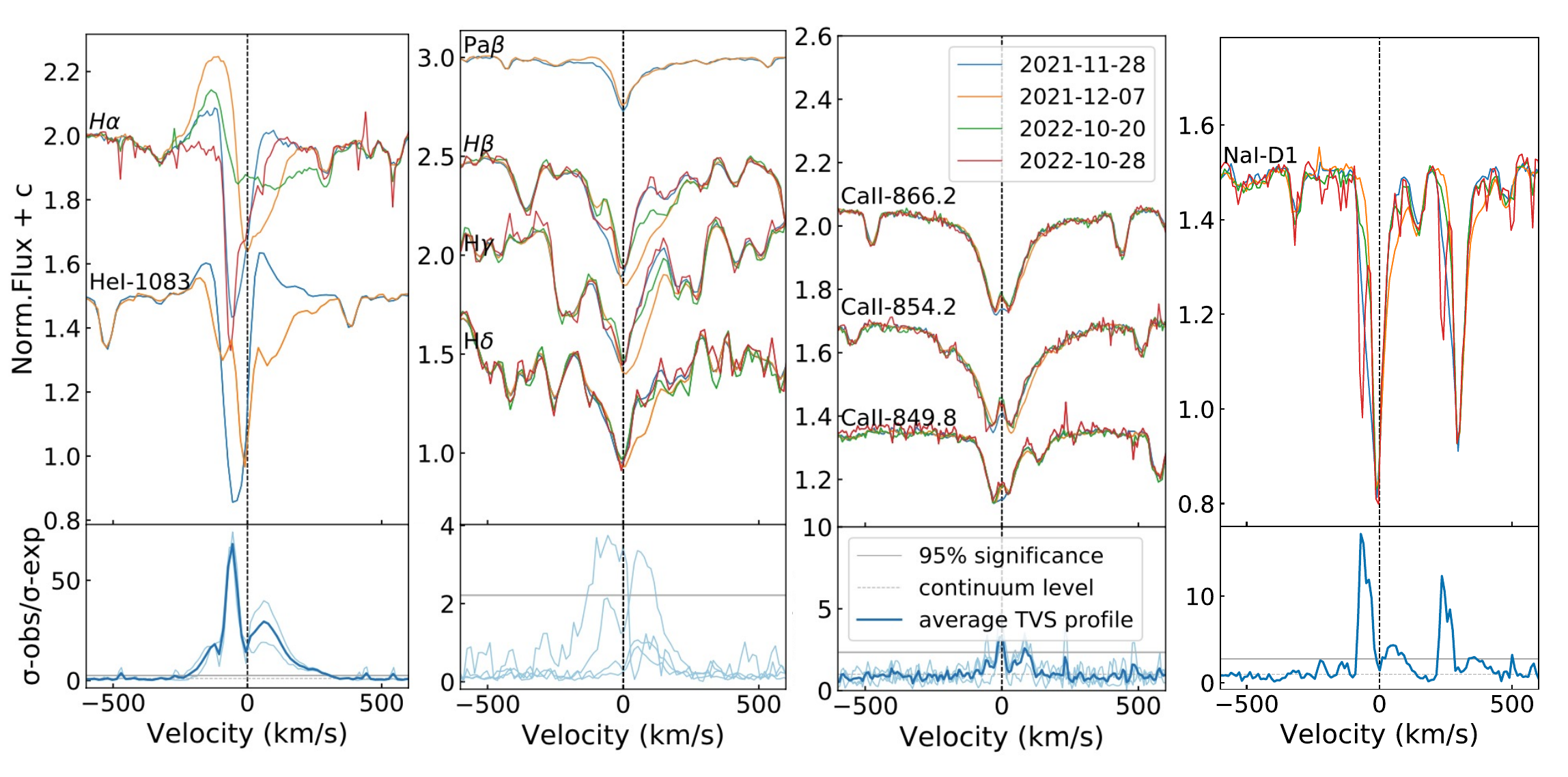}
\caption[]{Variable lines in PDS~111. From left to right: H$\alpha$ and He\,{\sc i} $\lambda$\,1083.0\,nm lines in the spectra of PDS\,111 are shown on velocity scale in the upper panel. They show significant variations. This is quantified in the lower panel by the use of the temporal variance spectrum, showing the significant variable velocity regime and similarity in the amplitude of the variation. The second panel shows three Balmer (H$\beta$, H$\gamma$ and H$\delta$) lines and Pa$\beta$, all displaying variations. The variations are quantified in the lower plot of that panel, showing variations in a similar velocity range as in the lower left panel. The third panel shows the Ca\,{\sc ii} triplet, and the last panel the Na\,{\sc i} doublet. The He\,{\sc i} $\lambda$\,1083.0\,nm and Pa$\beta$ line only include two spectra due to the limited wavelength range of HERMES compared to X-shooter. The HERMES observation on 28-10-2022 shows strong telluric contamination.
}
\label{fig: tvs}
\end{figure*}

The resulting TVS for all significantly varying lines are shown in Fig.~\ref{fig: tvs} and were used to measure the maximum velocity at which significant variability is detected. The lines that vary in a similar manner are shown together. The highest velocities and strongest variations are seen in H$\alpha$ and He\,{\sc i}, up to 320\,km\,s$^{-1}$ on the red side. The blue side shows variability up to 220\,km\,s$^{-1}$ which is similar to the highest variable velocities on both sides of the other H\,{\sc i} lines. Na\,{\sc i} lines show variations up to 120\,km\,s$^{-1}$ on the red side and up to 90\,km\,s$^{-1}$ on the blue side. The latter is similar to that seen in the red side of the components of the Ca\,{\sc ii} triplet. 

\begin{figure}
\centering
\includegraphics[width=\hsize]{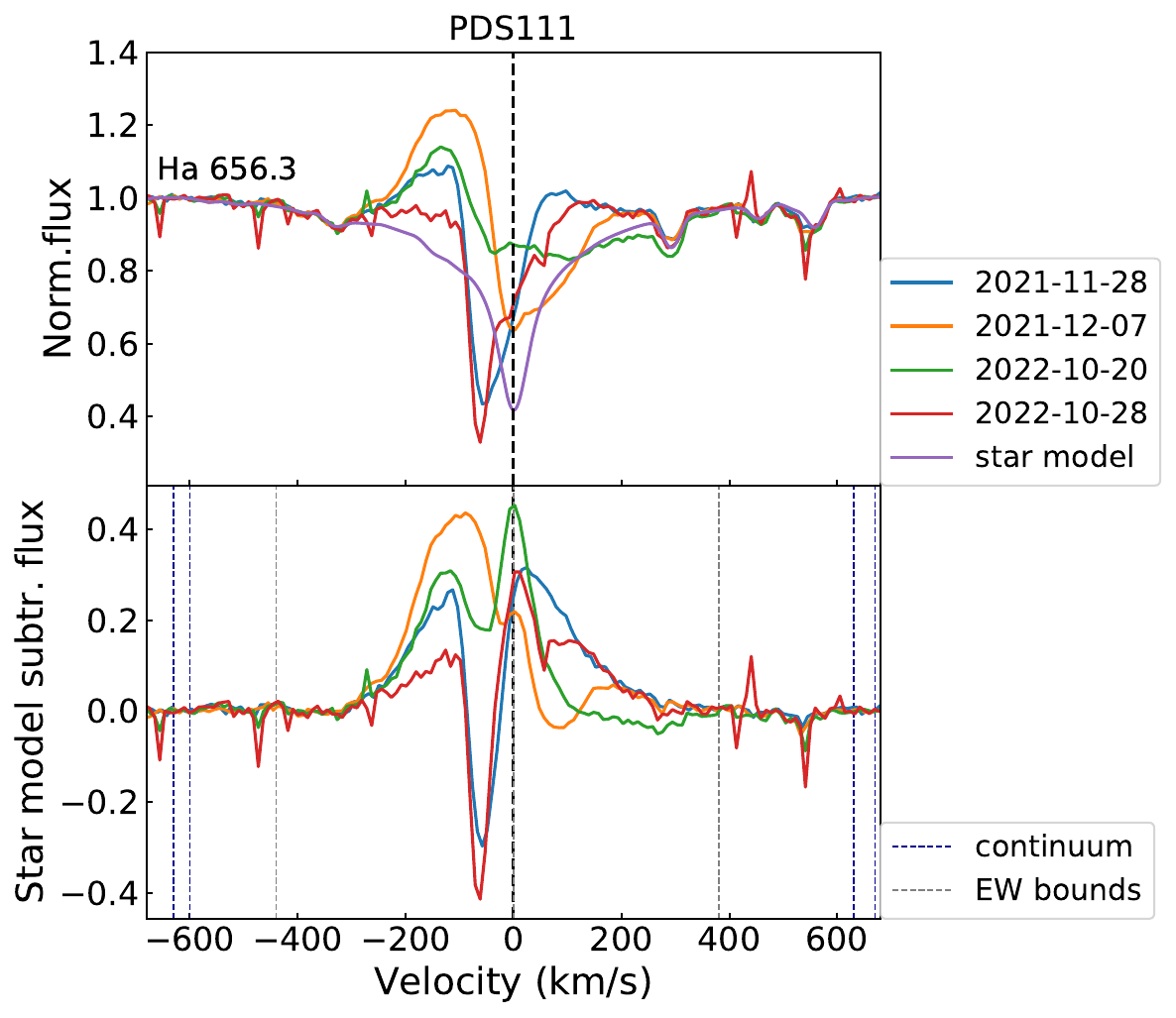}
\caption[]{The H$\alpha$ line of PDS\,111. The upper panel shows the H$\alpha$ lines of the four epochs and a star model. This star model is used to subtract the stellar contribution to H$\alpha$. The result is shown in the lower panel. The blue and gray dotted lines are the intervals used for normalization and measuring the equivalent width.}
\label{fig: Ha}
\end{figure}

The circumstellar H$\alpha$ emission, after star model subtraction and radial-velocity correction, has a strongly varying line profile (Fig.~\ref{fig: Ha}). The first and last epochs are characterized by a strong and sharp absorption component around $\sim$55\,km\,s$^{-1}$ on the blue side of the feature. The second epoch shows an inverse P\,Cygni profile and the third is characterized by two emission peaks, one around the rest wavelength and the other blue shifted. These profiles are discussed in Sec.\,\ref{p2:dis:varia}. 

\subsection{Binarity}
\begin{figure}[htbp]
\centering
\includegraphics[width=\hsize]{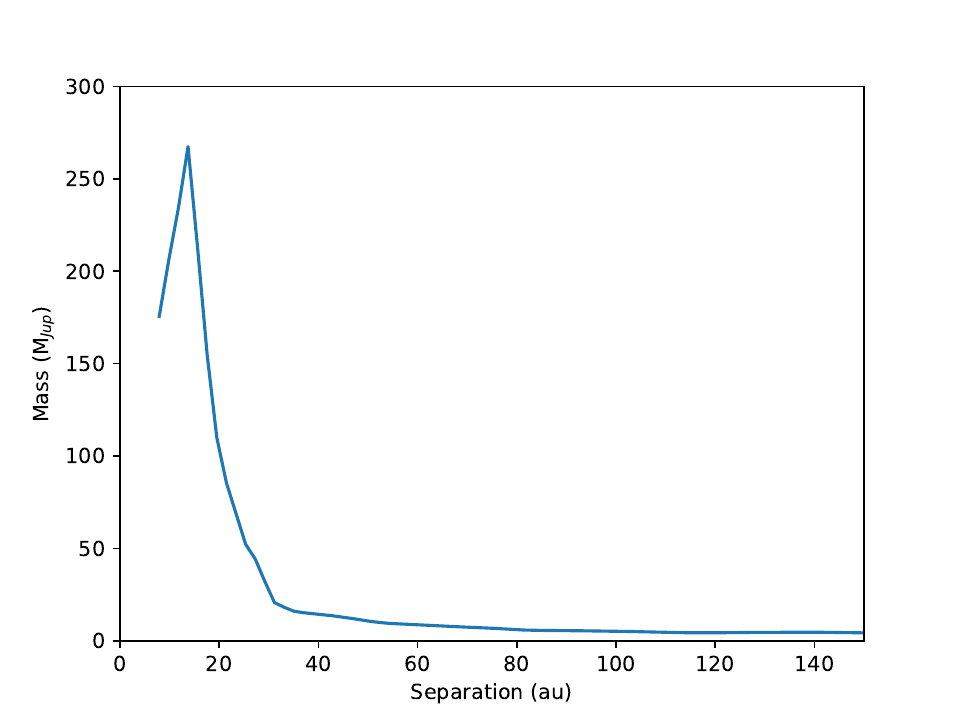}
\caption[]{Detectable minimum mass of co-eval (sub)stellar companions to PDS\,111 based on our SPHERE data processed with angular differential imaging.}
\label{fig: app: mass-detection}
\end{figure}

We calculated the contrast curves for the angular differential imaging final imaging. Those have been calculated using the Principal Component Analysis method. Then we assumed an age for the system of 16\,Myr and a distance of 158\,pc. The contrast curves were converted in mass limits of putative companions around the object. The mass limits are shown in Fig.~\ref{fig: app: mass-detection}. 

We observe no obvious detection of an inner binary component in the SPHERE data. Specifically, we can rule out objects down to $\sim$0.25\,M$_\odot$ and down to the coronagraphic edge at 0.1'', that is 16\,au. 

With these limits, we calculate the maximum expected radial velocity shift in the spectrum of the primary for a secondary mass of 0.25\,M$_{\odot}$. At 16\,au the radial velocity of the primary would be $\sim$1\,km\,s$^{-1}$, which would increase to $\sim$2\,km\,s$^{-1}$ when the distance between the primary and secondary is 8\,au and to $\sim$5\,km\,s$^{-1}$ for a separation of 1\,au). 

The mid-resolution X-shooter spectra and limiting signal-to-noise ratio of the HERMES spectra would be sufficient to detect radial velocity shifts down to at least 5\,km\,s$^{-1}$ if the epochs would be taken at opportune moments. This would mean that we should detect a secondary mass of 0.25\,M$_{\odot}$ if this star would have a separation <\,1\,au from the primary. The corresponding period of a few months at this separation is shorter than the time between the X-shooter and HERMES observations. Therefore, given our imaging and spectroscopic detection limits, the disk could potentially be a circumbinary disk, hiding a low mass secondary companion of a few tenths of M$_{\odot}$ at a separation between $\sim$1-16\,au.

\section{Discussion}\label{sec:discussion}

\subsection{A very flared, but old disk}
One of the most intriguing properties of PDS\,111 is the amount of flaring of the dust in the disk, which is comparable to systems that are much younger than PDS\,111 (Fig.~\ref{fig: scale-height}). If it is really this old, then such strong flaring suggests that vertical stirring of small grains is still efficient after $\sim$16\,Myr of evolution, implying that vertical turbulence remains at play. The origin and evolution of turbulence as well as the origin of small grains in protoplanetary disks is a matter of debate. Regarding turbulence, the traditional picture is that the magneto-rotational instabilities (MRI) would be the main source of disk turbulence, but the efficiency of MRI is a complex process that depends on dust properties and abundance, the magnetic field strength and configuration, the gas and dust chemistry, and various ionization sources \citep[see recent review in PPVII by][]{lesur2022}; most of these are not constrained by observations. Other hydrodynamical instabilities can contribute to vertical mixing, such as the vertical shearing instability \citep[e.g.,][]{flores2020, barraza2021}. In PDS\,111, whatever is dominating the dust vertical mixing still needs to be efficient after $\sim$16\,Myr of evolution, and be comparable to much younger disks, such as the flared disk of IM\,Lup \citep{Avenhaus2018}. \cite{Franceschi2023} model this star and suggest that the flaring could be caused by small particles that do not settle, because they do not grow after an initial growth time scale; some remain at high altitude because of long collisional time scales.

An alternative explanation is that the disk around PDS\,111 lacks turbulence and that the scattered light observations trace very small fluffy particles, which would not settle in the absence of vertical turbulence \citep{dullemond2004, kataoka2013}. Such particles can remain in the surface layers of the disk for several Myr if they do not grow and do not become compact. 

Another possible explanation is that the flaring is caused by small dust particles that are added to the disk surface by continuing infall on the disk surface or by blowing dust from the inner system to the disk surface. The possibility of infall from the surroundings to the disk \citep{dominik2008} is explored in Sec.~\ref{environment}, where we discuss the near environment of PDS\,111. 

We exclude a disk wind from causing the flared surface layer, as seen in RY\,Tau \citep{valegard2022}. A wind would make the surface of the disk more smooth without observable substructures in SPHERE, which is not the case in PDS\,111. We do observe hints of sporadic outflows (see Sec.\,\ref{p2:dis:varia}), but we do not observe any forbidden emission lines in the spectrum often related to a disk wind \citep{Nisini2018}. 

\subsection{Origin of the observed millimeter-cavity} \label{dis: cavity}

PDS\,111 displays a shallow cavity in the dust continuum emission that peaks at 30\,au, and which is not detected in scattered light or in the $^{12}$CO emission because the emission is hot and optically thick, similar to what has been seen in larger samples of transition disks \citep[e.g.,][]{villenave2019}. This cavity can originate from planet-disk interaction or (magneto-) hydrodynamical instabilities. In the case of planet-disk interaction, the planet mass should be $\lesssim$ M$_{\rm{Jup}}$ because otherwise a cavity would have been opened in small grains and that would leave a signature in scattered light \citep[][]{ovelar2013}, and potentially in the $^{12}$CO \citep{Facchini2018}. The mass of such a planet would allow gas to flow through the gap maintaining mass-accretion onto the star \citep{zhu2011, zhang2018}, which is moderate in PDS\,111 (Tab.~\ref{tab:Macc}). The detection of one or more planets is challenging with current telescopes \citep[e.g., with JWST,][]{wagner2024} because of their potential location ($\lesssim$30\,au) and low mass.

Alternatively, stellar binarity could also explain the presence of a cavity. Its shape of depends on the mass ratio, eccentricity, and disk viscosity \citep[e.g.,][]{artymowicz1994, miranda2017}. The location of the ring in the binary case can be 3.5-4 times the binary separation. For PDS\,111 that would be a binary separation around 8\,au. The binary scenario is further discussed in Sec.~\ref{dis:binary}.

Other origins of the cavity such as dead zones \citep[e.g.,][]{flock2015, delage2022} cannot be excluded. More observational constraints, such as disk kinematics, may help to understand its origin.

\subsection{Binarity}\label{dis:binary}
A circumbinary disk might explain several of the observed disk features such as the scattered light asymmetry, as well as the large disk cavity in the mm-continuum, as discussed in Sec.~\ref{dis: cavity}, for which the predicted separation is around $\sim$8\,au. Binarity could also explain the high RUWE value. \citet{juhasz2017} performed 3D SPH hydrodynamic simulations of a disk warp induced by a central binary. The synthetic scattered light images show that a disk warp can create strong illumination asymmetries, due to disk shadowing, especially when the warp "axis" is aligned with the disk minor axis. However, we note that their Fig.~5 (panel p) shows that for a continuous warp we would expect the illumination pattern to be reversed between disk top and bottom side. As we discuss in Sec.~\ref{sec:disk asymmetry}, this does not appear to be the case in PDS\,111, which may indicate a local increase in scale height rather than a continuous warp. 

Recent SPH hydrodynamic models by \cite{Ragusa2020}, \cite{munoz2020} and \cite{penzlin2022} suggest that a circumbinary disk that is co-planar with the orbital plane will always develop an eccentricity even if the initial binary orbit was close to circular. The main condition that they find is that the mass ratio between the binary components is $q>0.05$. This means in the case of PDS\,111 a binary component more massive than 55\,M$_\mathrm{Jup}$ would be required. This is not ruled out by our detection limits. Following the relation between cavity size and binary orbit recovered by \cite{Ragusa2020}, we would expect the binary component to be located at a semi-major axis of $\sim$8.5\,au. This is again consistent with our detection limits. However, we note that the presence of such a massive perturber at this relatively large separation might be in tension with our non-detection of the disk cavity in small grains, that is scattered light. 

\cite{Ragusa2020} find that their disks become increasingly eccentric because of resonant excitation or periodic "pumping" via oblique spiral shocks \citep{ogilvie2008, shi2012}. Due to the eccentricity, a "traffic jam" makes the density of the gas and dust higher at the apocenter than at pericenter \citep{ataiee2013, Ragusa2020}, potentially explaining the observed asymmetries in PDS\,111, as well as the possible local puffed up disk in scattered light. 

The ALMA data puts an upper limit on a potential cavity gap size caused by a stellar companion. If there is a gas cavity, we would have detected this in the gas emission if the cavity was larger than the beam size of the $^{12}$CO measurement, like in CS\,Cha \citep{kurtovic2022}. Therefore, the cavity must be smaller than 0.429\arcsec.

Literature only shows a few confirmed cases of prominent young, circumbinary disks seen with SPHERE, for instance HD\,142527 (\citealt{Fukagawa2006,Biller2012,Close2014}), HD\,34700 (\citealt{Monnier2019}) GG\,Tau (\citealt{Roddier1996, Keppler2020}), and CS\,Cha (\citealt{Ginski2018}). In most of these cases, the resolved scattered light disk shows strong azimuthal asymmetries like spiral arms, which are not in our observations of PDS\,111. However, this may strongly depend on the binary separation, eccentricity, mass ratio, and orbit inclination relative to the disk. We note that in the case of the CS\,Cha circumbinary disk the outer scattered light disk appears completely smooth and featureless, highlighting the diversity in disk morphology. 

\subsection{Environment}\label{environment}

PDS\,111 is known to be a foreground star of the Orion cloud \citep{Torres1995}, recently also confirmed by its {\it Gaia} DR3 distance \citep[$\varpi\, =\, 6.3136\pm0.0352$\,mas;][]{Gaia2022}. No strong arguments are found that suggest the star is a member of a young cluster. PDS\,111 has been associated with the Columba association \citep{Torres2008, dasilva2009, Launhardt2022}. However this predates the availability of an accurate trigonometric parallax measurement 
\citep[Tycho parallax was $\varpi\, =\, 28.70\pm21.60$\,mas;][]{hipparcos1997}. Columba is, however, older \citep[$42^{+6}_{-4}$\,Myr;][]{Bell2015} and an early G-type member would already be on the
main sequence. The amount of circumstellar matter and our evolutionary age imply PDS\,111 is considerably younger. Another membership assignment from the literature is with Theia 160 \citep{Kounkel2019}, a clumpy string of co-distant and co-moving stars associated with PDS\,111. This string shows young stars, but also includes stars that have been attributed to a much older (62$\pm$7\,Myr) group, the $\mu$ Tau association \citep{Gagne2020}. More investigation into the young stars associated with Theia 160 could lead to a PMS track age, which would give additional constraints on the age of the star. 
So, its genetic connection to any characterized young stellar groups is uncertain at present. 

Furthermore, its environment could be important for a rejuvenation scenario, where streamers feed material to the disk \citep{Kuffmeier2020, Ginski2021}. Arc-like structures, spiral arms, disk misalignment, or other disk structures in later stages of star formation, might be related to late infall of material to the disk. Therefore, late infall could have influenced the flaring and asymmetry in the disk of PDS\,111.   
The STILISM 3D maps \citep{lallement2018} predict reddening toward PDS\,111 at distance of 158\,pc to be very modest E($B$-$V$) = $0.008\pm0.016$\,mag ($A_V$ = $0.025\pm0.050$ mag), and the reddening is not expected to exceed E($B$-$V$) $\sim$ 0.05\,mag until $\sim$475\,pc (i.e., near Orion complex), so there does not seem to be any dense, high extinction interstellar dust clouds in its vicinity. Given its disk, any significant reddening or changes in reddening must be due to its own circumstellar disk. Therefore, it seems unlikely that the disk is rejuvenated by its local environment. 

\subsection{H$\alpha$ and He\,{\sc i} variability}\label{p2:dis:varia}
The hot gas from which H$\alpha$ originates is highly variable in line morphology and shows velocities up to 320\,km\,s$^{-1}$. These velocities, assuming a Keplerian rotating disk, are formed close to the stellar surface. The inverse P\,Cygni profile, seen in the epoch of 2021-12-07, is widely interpreted as a signature of accretion \citep[e.g.,][]{cauley2014} and has a red-shifted absorption component in the feature around $\sim$100\,km\,s$^{-1}$. The deep narrow blue-shifted absorption component in the first and last epoch is strongly variable 
and resides at relatively low velocities \citep[$< 100$\,km\,s$^{-1}$,][]{Nisini2018, bouvier2023}. This He\,{\sc i} feature shows a similar blue-shifted absorption feature in the one epoch that also has an X-shooter observation. Since the blue-shifted absorption is deep, narrow and variable, it probably arises when some cooler optically thick material moves in front of the hot gas from the inner disk where the H$\alpha$ and He\,{\sc i} lines are formed. These blue-shifted features are generally associated with outflows \citep{kwan2007, Nisini2018, bouvier2023}.

\cite{cauley2014} observed similar profiles in a large sample of He\,{\sc i} emission lines from PMS stars and imply an origin of the line morphology from a misalignment between the inner and outer disk. \cite{bouvier2023} also showed a similar line morphology of the He\,{\sc i} and Balmer lines in the classical T\,Tauri star GM\,Aur with a mass accretion rate of $\sim$10$^{-8}$\,M$_{\odot}$\,yr$^{-1}$ and an inclination of $\sim$68$^{\circ}$. By combining optical and NIR photometry, and high-resolution spectroscopic time series, they concluded that these line features are caused by sporadic mass loss from the disk surface. The blue-shifted narrow absorption components in GM\,Aur start at high velocities, move to lower velocity over a period of days and disappear. Whether PDS\,111 shows similar mass loss or an inner-disk misalignment is to be confirmed by follow-up observations with a higher time cadence (e.g., daily).

\subsection{Long-lived accretion disks}\label{p2:sec:peterpan}
Recently, \cite{Silverberg2020} found a class of long-lived accretion disks around M-dwarfs with ages above 20\,Myr, so-called "Peter Pan disks" for their refusal to "grow up" and disperse. They concluded that these disks were most likely gas-poor hybrid disks, exhibiting signs of young and debris disks, but could not explain why such old systems still harbored these disks. \cite{Coleman2020} used disk evolutionary models to study possible conditions for a gas-rich disk to survive around these old stars. They concluded that a low turbulence parameter $\alpha\sim$10$^{-4}$, a high initial disk mass ($>$0.25\,M$_*$) and very low external photo-evaporation ($<$10$^{-9}$\,$\mathrm{M_\odot \, yr^{-1}}$) can reproduce the "Peter Pan disks" around M-dwarfs. Importantly, they also studied how these conditions would change for higher central star masses, as would be the case for the PDS\,111 system. They conclude that for a solar mass star, a disk can be maintained up to 50\,Myr under similar conditions, if external photo-evaporation is extremely low ($<$10$^{-10}$\,$\mathrm{M_\odot\,yr^{-1}}$).

\subsection{A coherent picture of the disk of PDS\,111}
We speculate that both radial and azimuthal sub-structures could be caused by an internal perturber. Either a low-mass planet close to the cavity edge at $\sim$30\,au, which may explain the continuum cavity as well as the absence of such a cavity in the gas or scattered light, or a close-in stellar binary located between 1.5 and 16\,au, with a preference for 8.5\,au to explain the mm cavity size. Such a binary may then also explain the azimuthal asymmetry by exciting disk eccentricity.
The presence of a perturber in the disk cavity might also go some way to explain the presence of a gas-rich disk at the old age of the system by trapping material in the outer disk. This may be consistent with the low accretion rate of 1-5$\times$10$^{-10}$\,M$_{\odot}$\,yr$^{-1}$ that we inferred from our H$\alpha$ and H$\beta$ measurements. Still, planet formation and thus cavity opening is presumably a common occurrence in young disks. Related sub-structures are indeed detected in numerous systems with high-resolution ALMA or scattered light observations (e.g., \citealt{andrews2018, benisty2022}). Yet studies that measure the disk occurrence rate in young clusters typically find that most disks have dispersed after 10\,Myr (e.g., \citealt{Mamajek2009, 2010A&A...510A..72F, Ribas2014}).

\section{Summary and Conclusions}
\label{sec: Sum}

We present the results of an extensive observational campaign of the hitherto relatively unexplored PDS\,111 system in the framework of the DESTINYS program. Our spectroscopic, imaging and mm-interferometry observations explore the system from the stellar photosphere out to the edge of the gaseous disk at $\sim$295\,au. They reveal a system that appears archetypal in many ways for a young gas-rich circumstellar disk. Yet our spectroscopic analysis of the central star as well as the possible membership in the Columba association place the system age at an age of $\sim$16\,Myr (for spectroscopy) or even significantly older (for association membership). Equally puzzling is the relative isolation of PDS\,111 in the foreground of the Orion star-forming region. 

Our NIR scattered light observations as well as ALMA continuum observations resolve the outer disk and reveal radial sub-structures as well as azimuthal asymmetries. The azimuthal asymmetry might trace a warp or local overdensity in the outer disk. Interestingly, the complex and variable profiles of the H$\alpha$ line might also indicate a misalignment of the inner disk region.

The "Peter Pan disk" scenario for higher mass central stars fits well with the relative isolation of PDS\,111. Thus it is possible that the combination of low external photo-evaporation due to lack of close association neighbors as well as the cavity opening and subsequent low viscous transport of material within the disk conspired in this system to maintain a young-looking disk around an older star. Thus PDS\,111 might well be the first high mass equivalent of the "Peter Pan" population among M-dwarfs. Our extensive data set, and in particular our spatially resolved observations of the disk, make PDS\,111 the ideal laboratory to study this class of long-lasting disks with dedicated simulations. 

\begin{acknowledgements}
SPHERE was designed and built by a consortium made of IPAG (Grenoble, France), MPIA (Heidelberg, Germany), LAM (Marseille, France), LESIA (Paris, France), Laboratoire Lagrange (Nice, France), INAF–Osservatorio di Padova (Italy), Observatoire de Genève (Switzerland), ETH Zurich (Switzerland), NOVA (Netherlands), ONERA (France) and ASTRON (Netherlands) in collaboration with ESO. SPHERE was funded by ESO, with additional contributions from CNRS (France), MPIA (Germany), INAF (Italy), FINES (Switzerland) and NOVA (Netherlands).

Additional funding from EC's 6th and 7th Framework Programmes as part of OPTICON was received (grant number RII3-Ct-2004-001566 for FP6 (2004–2008); 226604 for FP7 (2009–2012); 312430 for FP7 (2013–2016)). We acknowledge the Programme National de Planétologie (PNP) and the Programme National de Physique Stellaire (PNPS) of CNRS-INSU, France, the French Labex OSUG@2020 (Investissements d’avenir – ANR10 LABX56) and LIO (Lyon Institute of Origins, ANR-10-LABX-0066 within the programme Investissements d'Avenir, ANR-11-IDEX-0007), and the Agence Nationale de la Recherche (ANR-14-CE33-0018) for support. A.Z. acknowledges support from ANID -- Millennium Science Initiative Program -- Center Code NCN2021\_080.

This paper makes use of the following ALMA data: ADS/JAO.ALMA\#2021.1.01705.S. ALMA is a partnership of ESO (representing its member states), NSF (USA) and NINS (Japan), together with NRC (Canada), MOST and ASIAA (Taiwan), and KASI (Republic of Korea), in cooperation with the Republic of Chile. The Joint ALMA Observatory is operated by ESO, AUI/NRAO and NAOJ.

Based on observations obtained with the HERMES spectrograph, which is supported by the Research Foundation - Flanders (FWO), Belgium, the Research Council of KU Leuven, Belgium, the Fonds National de la Recherche Scientifique (F.R.S.-FNRS), Belgium, the Royal Observatory of Belgium, the Observatoire de Genève, Switzerland and the Thüringer Landessternwarte Tautenburg, Germany.

This paper includes data collected by the TESS mission. Funding for the TESS mission is provided by the NASA's Science Mission Directorate.

P.P. acknowledge funding from the UK Research and Innovation (UKRI) under the UK government’s Horizon Europe funding guarantee from ERC (grant agreement No 101076489)

This project has received funding from the European Research Council (ERC) under the European Union's Horizon Europe research and innovation program (DUST2PLANETS,  grant agreement No. 101053020).

This project has received funding from the European Research Council (ERC) under the European Union’s Horizon 2020 research and innovation programme (PROTOPLANETS, grant agreement No. 101002188). 

Part of this research was carried out at the Jet Propulsion Laboratory, California Institute of Technology, under a contract with the National Aeronautics and Space Administration (80NM0018D0004).

\end{acknowledgements}

\bibliographystyle{aa} 
\bibliography{references} 


\begin{appendix}
\onecolumn

\section{Flux calibrated stacked total intensity and polarimetric images}
\begin{figure*}[ht]
\centering
\includegraphics[width=0.9\textwidth]{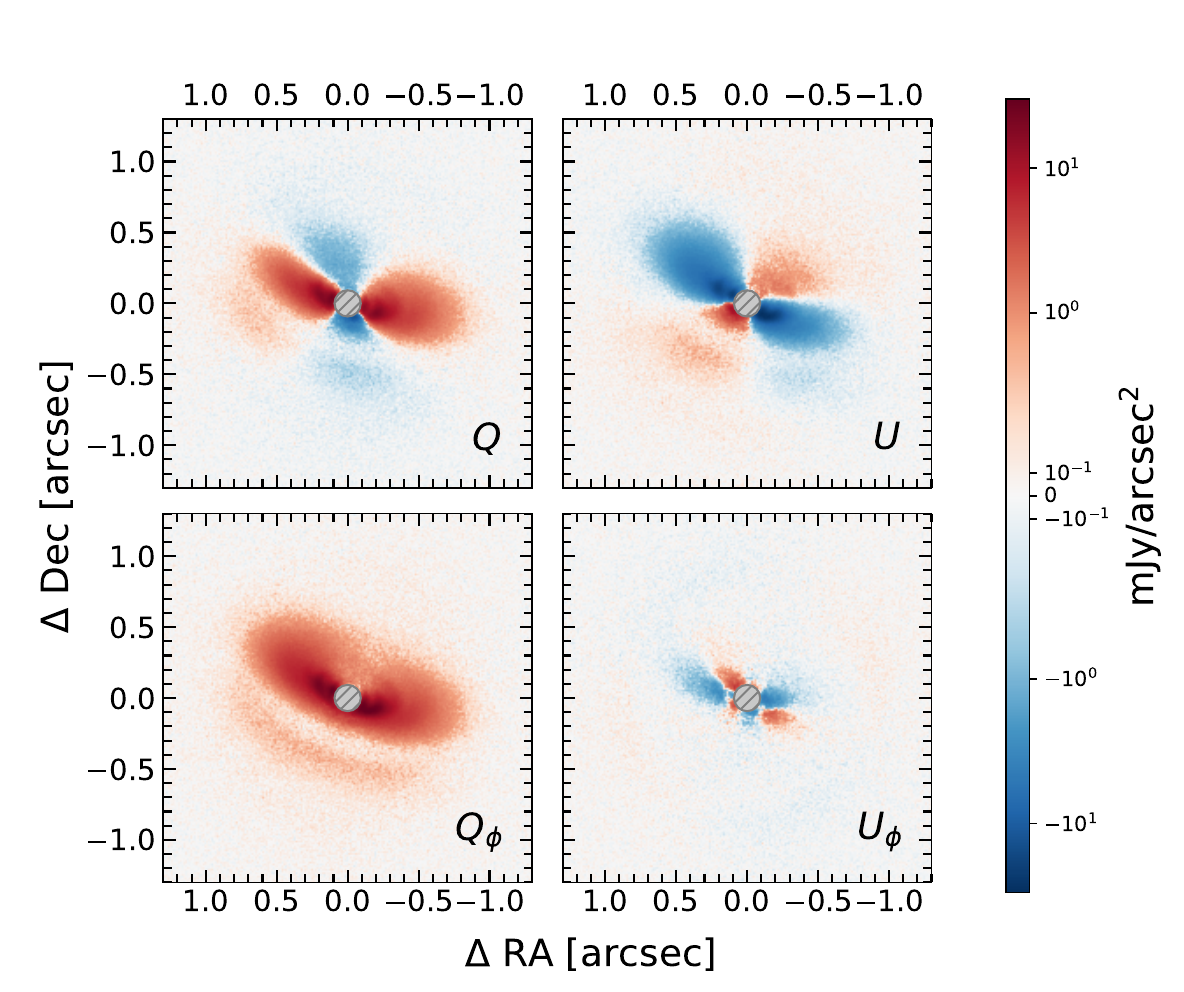}
\caption[]{Flux calibrated polarimetric images. The directly measured Stokes $Q$ and $U$ parameters are shown in the top row and the derived azimuthal Stokes $Q_\phi$ and $U_\phi$ parameters in the bottom row. All images are shown on the same (logarithmic) scale, symmetric around 0. In all images, the gray, hashed circle in the center marks the area that was covered by a coronagraphic mask.}
\label{fig: app: flux-pol}
\end{figure*}

\begin{figure}[ht]
\centering
\includegraphics[width=0.8\textwidth]{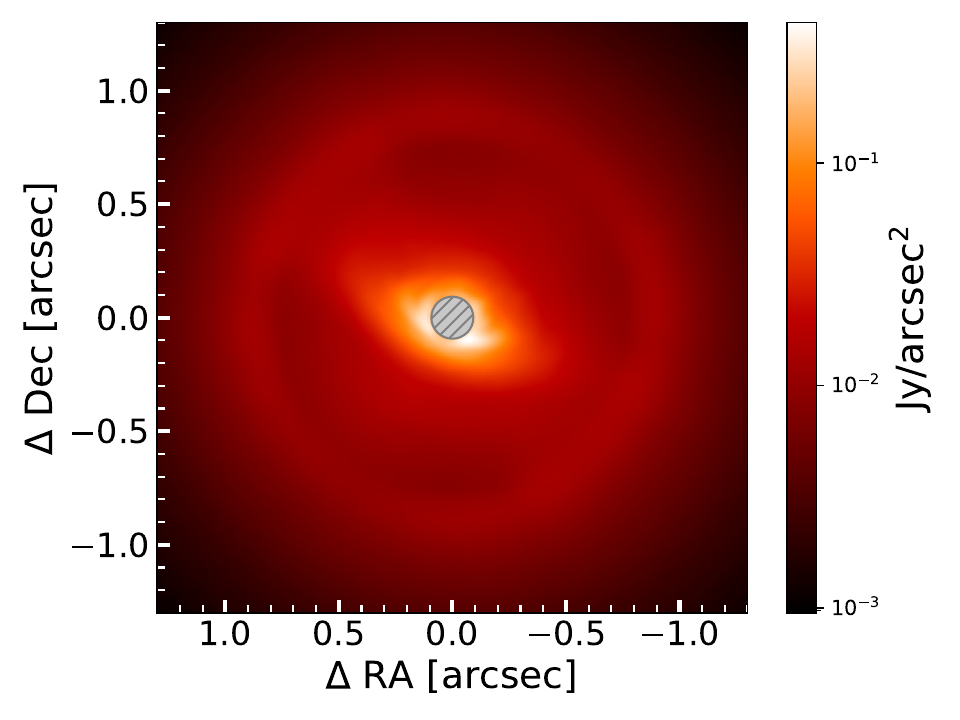}
\caption[]{Flux calibrated stacked total intensity image. We note that in this image the stellar light was not removed. The bright ring visible on the outer edge of the image shows the adaptive optics correction radius which shows bright stellar speckle signal (smoothed out by the field rotation during observations). The signal of the disk is faintly visible close to the coronagraph. The image is shown on a (logarithmic) scale, symmetric around 0. The grey, hashed circle in the center marks the area that was covered by a coronagraphic mask.}
\label{fig: app: stacked}
\end{figure}

In Fig.~\ref{fig: app: stacked} we show the de-rotated and stacked total intensity image, which contains the disk and stellar signal.
In Fig.~\ref{fig: app: flux-pol} we show the flux calibrated Stokes $Q$, $U$, $Q_\phi$, $U_\phi$ images. Flux calibration was done with the central star as the reference source. For this purpose we used the flux calibration frames which were taken before and after the science images and in which the star is removed from the coronagraph and a neutral density filter inserted to prevent saturation. We measured the central star flux using aperture photometry. We then translate detector counts into units of surface brightness via the known H-band magnitude of the primary star. 

\label{app:pol-flux}

\newpage
\section{Stellar atmosphere model fit}\label{p2:app:chi2}
\begin{figure*}[ht]
\centering
\includegraphics[width=0.6\hsize]{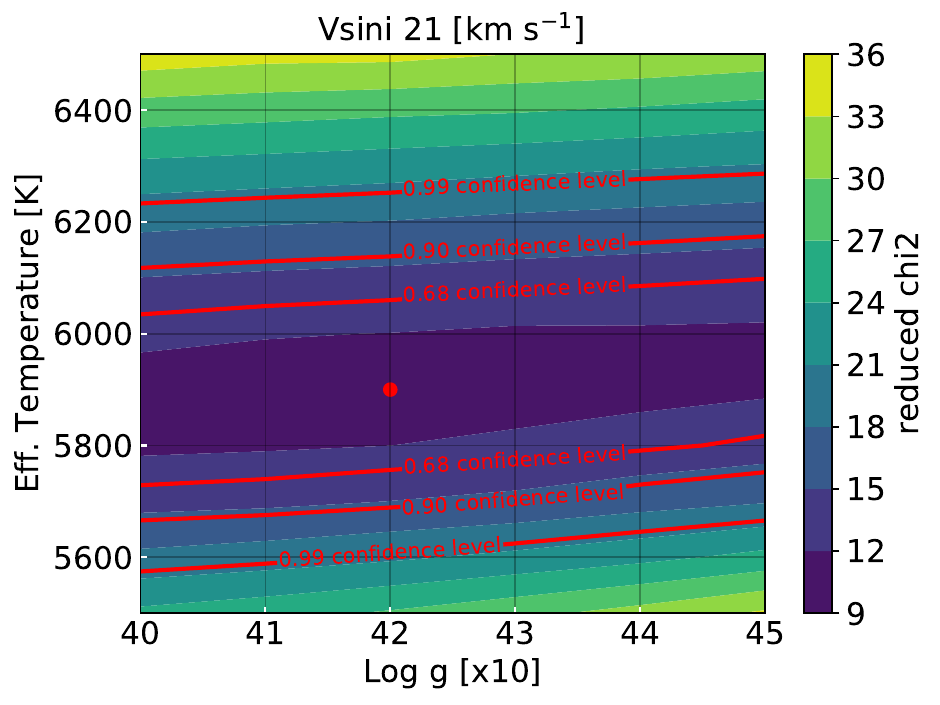}
\caption[]{Resulting reduced $\chi^{2}$ values from atmospheric model fitting. The Kurucz models have different temperatures and surface gravities and a constant $v$\,sin$i$. The lowest $\chi^{2}$ value is indicated with a red dot.}
\label{p2:fig:chi2fit}
\end{figure*}

The $\chi^{2}$ values from atmospheric model fitting are shown in Fig.\,\ref{p2:fig:chi2fit} and a subset of the models, including the best fit, is shown in Fig.\,\ref{p2:fig:models}. The fitting range is from 402\,nm to 430\,nm based on \cite{gray2000digital}.

\section{SED}\label{p2:app:sed}

\begin{figure}[ht]
\centering
\includegraphics[width=0.45\hsize]{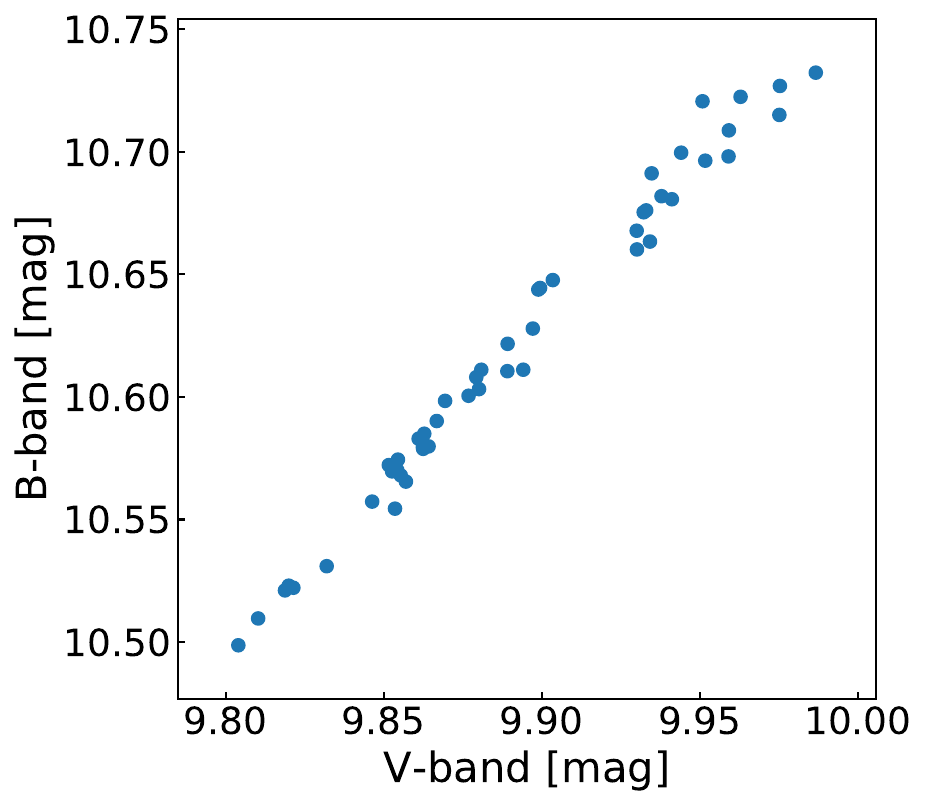}
\caption[]{B- and V-band photometry converted from Gaia measurements. The B- and V-band measurements show variations.}
\label{p2:fig:photometry1}
\end{figure}

\begin{figure}[ht]
\centering
\includegraphics[width=0.45\hsize]{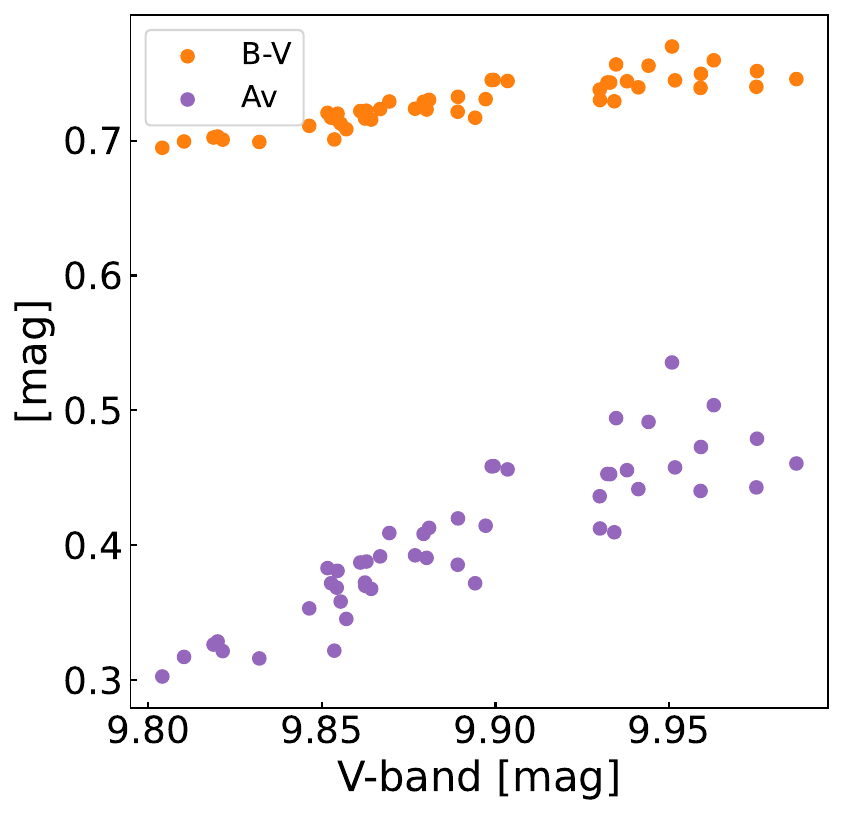}
\caption[]{$B$-$V$ color and $A_{\rm V}$ shown for V-band measurements determined from Gaia. $B$-$V$ shows a slight increase with increasing V-band magnitude. The reported $A_{\rm V}$ follows from $A_{\rm V} = R_{\rm V}* \text{E}(B\text{-}V)$ where E($B$-$V$) is determined by comparing the observed $B$-$V$ to the intrinsic $(B$-$V)_0$ of 0.597\,mag (interpolated from table 6 in \cite{Pecaut2013} for a star of 5900\,K) and also increases with an increasing V-band measurement. The $A_{\rm V}$ values are slightly higher than determined by the SED fit.}
\label{p2:fig:photometry2}
\end{figure}

\begin{figure*}[ht]
\centering
\includegraphics[width=0.6\hsize]{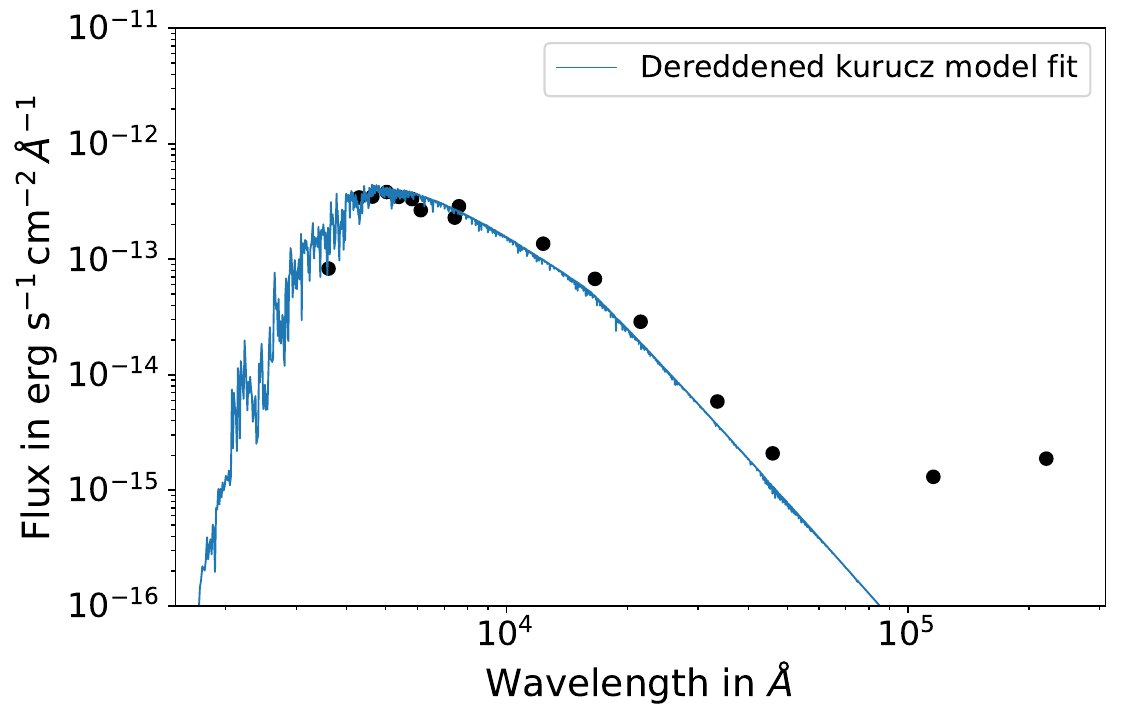}
\caption[]{The SED following from Kurucz model fitting to optical photometry, discussed in Sec.\,\ref{p2:sec:sed}. The model and photometry are dereddend using Cardelli's extinction law. Infrared excess is observed from the J-band on toward redder wavelengths.}
\label{p2:fig:sed}
\end{figure*}

In this section we show the B- and V-band from Gaia photometry and the SED discussed in Sec.~\ref{p2:sec:sed}. Figs.\,\ref{p2:fig:photometry1} and \ref{p2:fig:photometry2} shows the B- and V-band measurements converted from Gaia measurements \citep{Gaia2022, Jayasinghe2018}. Fig.\,\ref{p2:fig:sed} shows the SED and Table\,\ref{p2:tab:sed_phot}.

\begin{table}[]
\caption{Photometry used for SED fitting.}
\centering
\begin{tabular}{lllll}
\toprule
Band      & Flux [mag] & error [mag] & effective wavelength [\AA] & reference \\
\midrule \midrule
\textit{u} (SDSS)  &   11.638      & 0.028       & 3608 &  \cite{2018PASA...35...10W}     \\
\textit{B} (Johnson) & 10.70  & 0.038  &  4299  &  See Sec.\,\ref{sec:stellarprops}       \\
\textit{g} (SDSS)    & 10.506	   & 0.12	  & 4639     &   \cite{2018PASA...35...10W}     \\
\textit{G$_{\text{BP}}$} (Gaia) & 10.0731    & 0.00766 &  5036     &  \cite{Gaia2022}       \\
\textit{V} (Johnson) & 9.97  & 0.22   &  5394	&  See Sec.\,\ref{sec:stellarprops}      \\
\textit{G} Gaia   &   9.69303  & 0.00341 &  5822     &  \cite{Gaia2022}      \\
\textit{r} (SDSS)     &  9.943	   & 0.04	 &  6122     &  \cite{2018PASA...35...10W}     \\
\textit{i} (SDSS)  &   9.465	   & 0.05	 &  7439     &  \cite{2018PASA...35...10W}       \\
\textit{G$_{RP}$} (Gaia)     & 9.11425  & 0.00612 &  7620     &  \cite{Gaia2022}       \\
\textit{J} (2MASS)    & 8.403      & 0.02   & 12350	    &  \cite{2006AJ....131.1163S}     \\
\textit{H} (2MASS)    & 8.058      & 0.55   & 16620	    &  \cite{2006AJ....131.1163S}     \\
\textit{Ks} (2MASS)   & 7.932      & 0.024  & 21590	    &  \cite{2006AJ....131.1163S}     \\
\textit{W1} (WISE)    & 7.862      & 0.024  & 33526	    &  \cite{2010AJ....140.1868W}      \\
\textit{W2} (WISE)   & 7.662      & 0.02   & 46028	    &  \cite{2010AJ....140.1868W}       \\
\textit{W3} (WISE)  & 4.246      & 0.014  & 115608     &  \cite{2010AJ....140.1868W}      \\
\textit{W4} (WISE)   & 1.083      & 0.01   & 220883     &  \cite{2010AJ....140.1868W}      \\
\bottomrule
\end{tabular}
\label{p2:tab:sed_phot}
\end{table}


\newpage
\section{EW measurements}
The EW measurements of several lines are reported in Tab.\,\ref{tab:EW}.

\begin{table*}[ht]
\centering
\caption{EW measurements in \AA\ after stellar model subtraction. }
\begin{tabular}{ll | llll}
\toprule \midrule
               & $\lambda_{rest}$ (in nm)  & 28-11-2021 & 07-12-2021 & 20-10-2022 & 28-10-2022 \\
                 \midrule
H$\alpha$  &   656.279   & -1.274     & -1.873     & -1.567     & -0.811     \\
H$\beta$   &   486.135   & -0.240     & 0.198      & -0.071     & -0.239     \\
Li\,{\sc i} &  670.78    & 0.237      & 0.239      & 0.249      & 0.253      \\
Ca\,{\sc ii}-i &  849.802  & 0.099      & 0.078      & 0.052      & 0.031      \\
Ca\,{\sc ii}-ii & 854.209 & 0.411      & 0.406      & 0.249      & 0.311      \\
Ca\,{\sc ii}-iii &  866.214  & 0.142      & 0.155      & 0.111      & 0.167      \\
He\,{\sc i} &  1083.33 & 0.669      & 1.813      &            &       \\ 
\midrule  
\bottomrule
\end{tabular}

\label{tab:EW}
\end{table*}

\newpage
\section{Ellipse fitting}

\begin{figure*}[ht]
\centering
\includegraphics[width=0.999\textwidth]{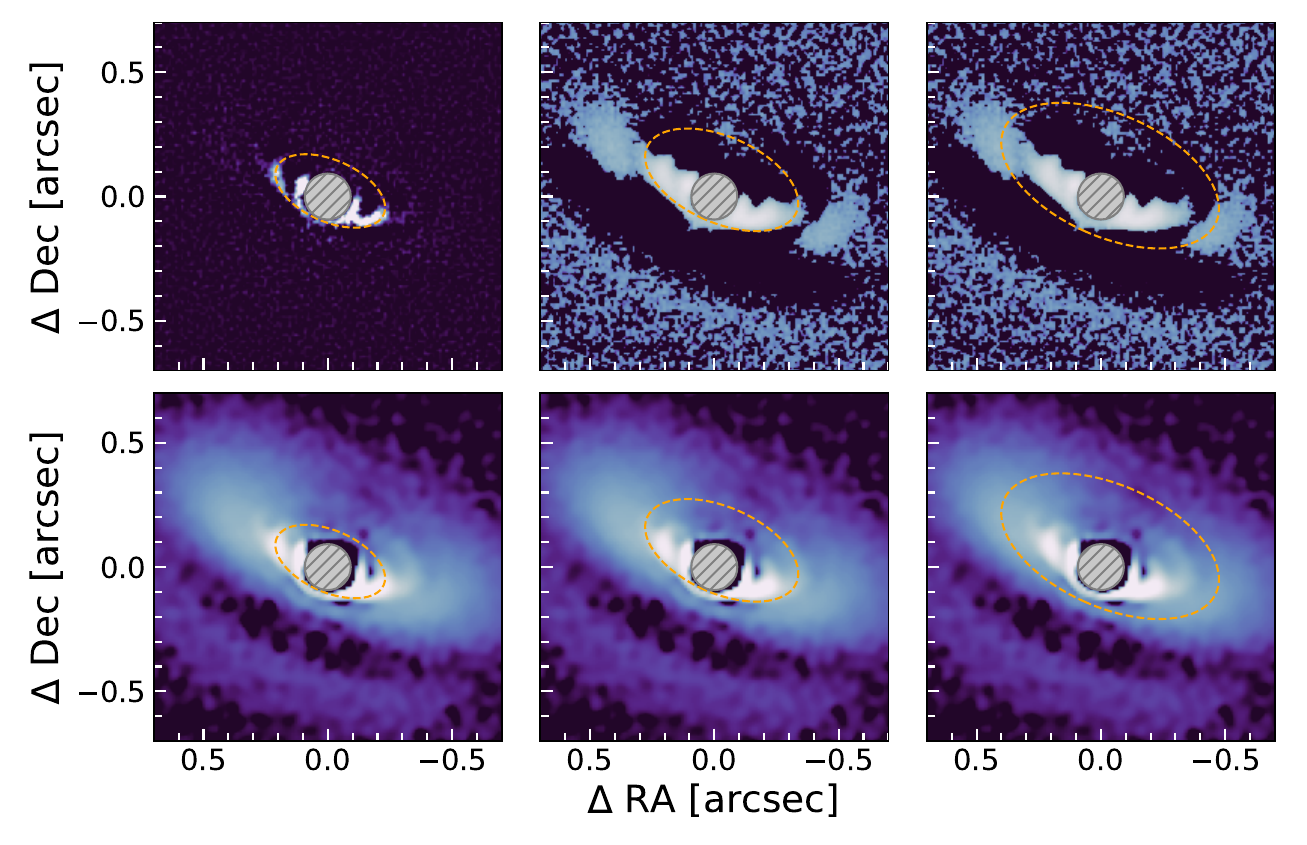}
\caption[]{SPHERE scattered light data after high pass filtering (top row) and deconvolution (bottom row). The color map was adjusted in each panel to highlight the most prominent sub-structure in the disk. We overplot ellipses tracing the two inner rings and the following gap from left to right.}
\label{fig: app: ellipse}
\end{figure*}

\begin{figure*}[htbp]
\centering
\includegraphics[width=0.59\textwidth]{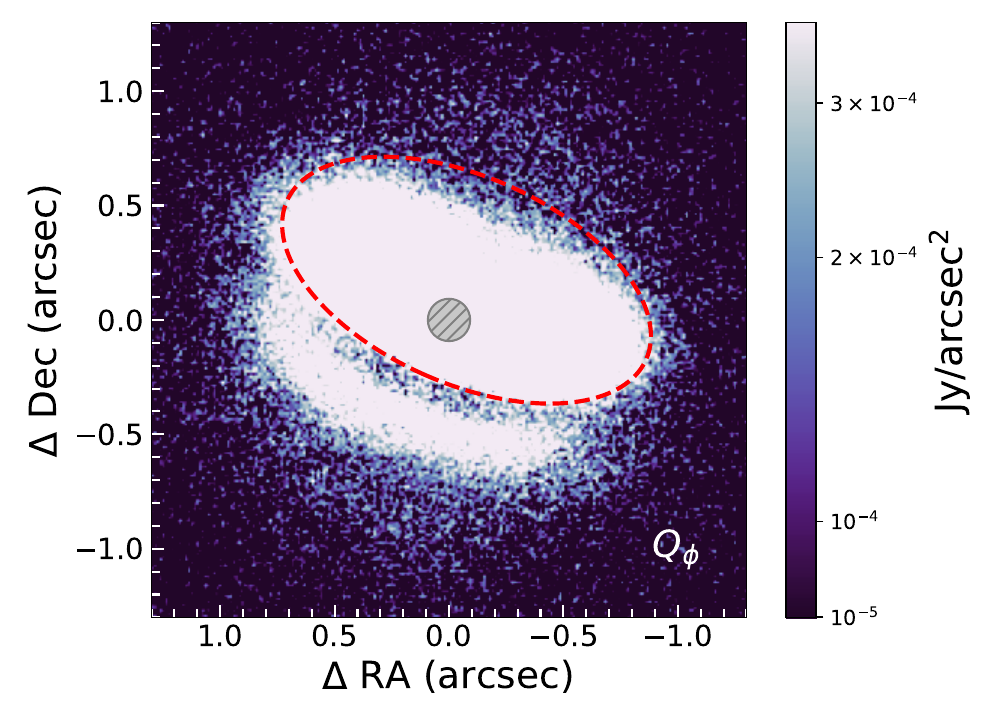}
\caption[]{SPHERE $Q_\phi$ image shown on a saturated color map to highlight the outer edge of the disk. The red dashed line marks the ellipse also shown in Fig.~\ref{fig: scale-height}, which traces the outer edge. Note that due to the scattering phase function the disk far side is faint, and thus appears only at low S/N.}
\label{fig: app: ellipse-saturated}
\end{figure*}

Fig.~\ref{fig: app: ellipse} and \ref{fig: app: ellipse-saturated} highlight the disk structures and the overlayed ellipses.

\label{app:ellipse}

\section{Components of the visibility fitting}

\begin{figure*}[ht]
 \centering
        \includegraphics[width=8.4cm]{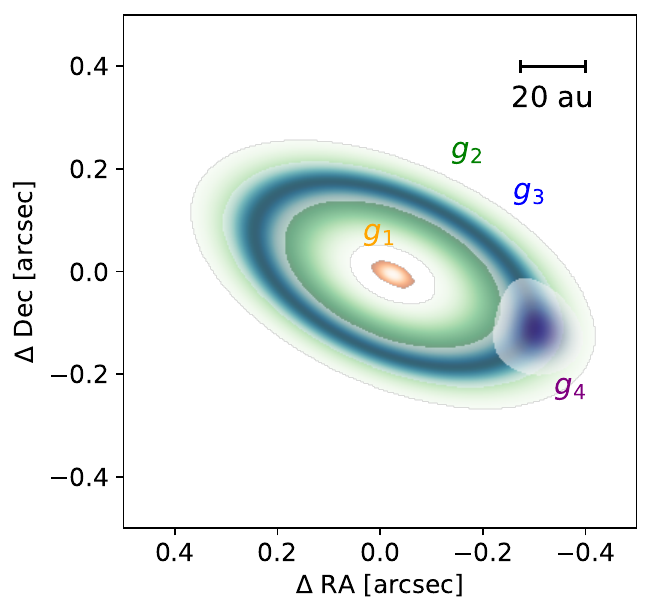}
   \caption{Schematic of PDS\,111 components of the model used to fit the visibilities from the ALMA observations. The model components do not reflect the relative flux of contribution. }
   \label{fig:components}
\end{figure*}

Fig.~\ref{fig:components} shows a sketch of the components of the intensity profile that are assumed for the visibility fitting of the dust continuum emission of the disk around PDS\,111 observed with ALMA.

\newpage
\section{Channel map of the $^{12}$CO}

\begin{figure*}[htbp]
\centering
\includegraphics[width=0.99\hsize]{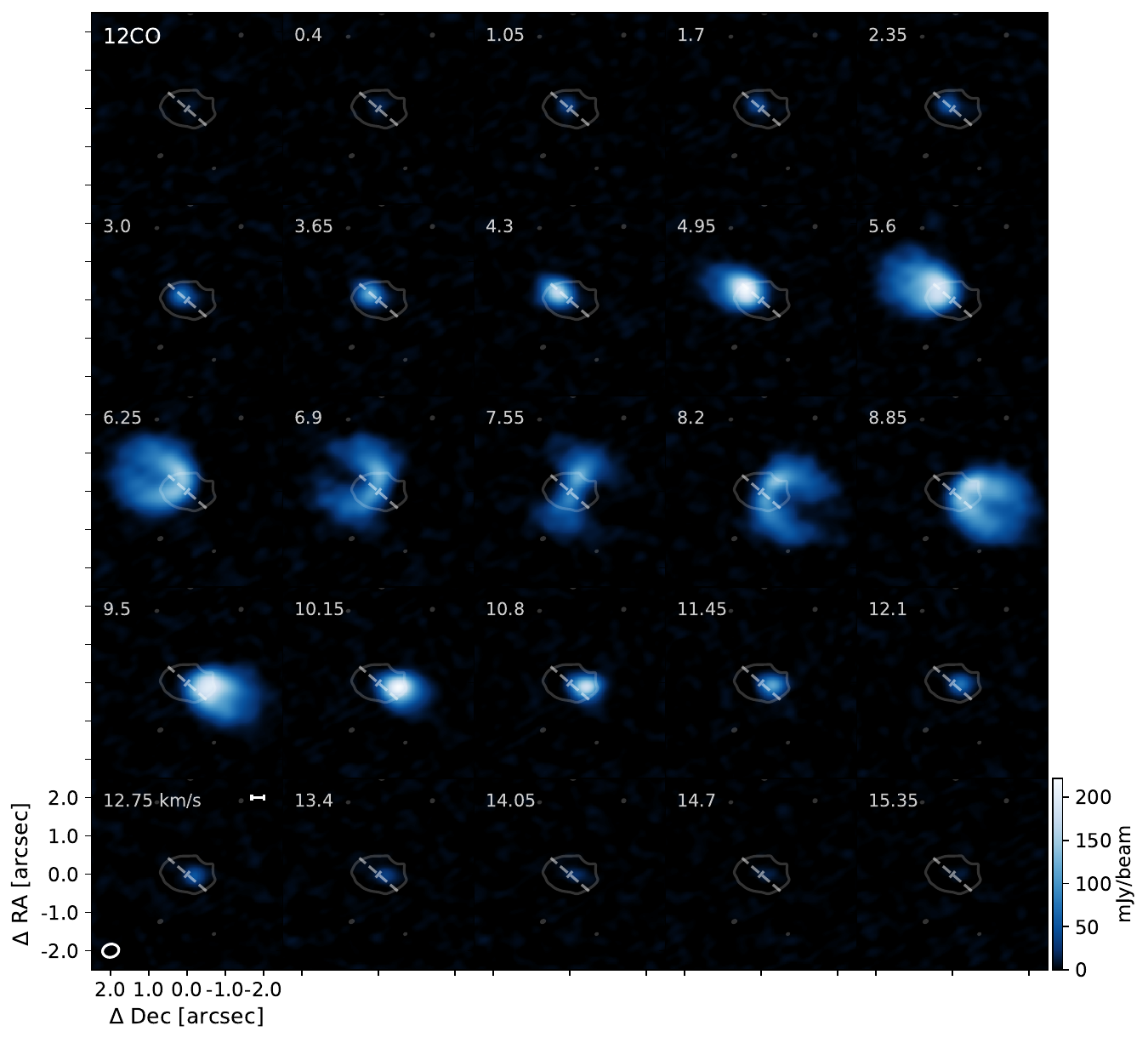}
\caption[]{Channel maps of $^{12}$CO of PDS\,111. The gray contours are at the $5\times\sigma$ level of the continuum emission. The scale bar in the left panel represents a scale of 50\,au. The dashed line is the PA of the continuum emission in the disk.}
\label{fig:channel_maps}
\end{figure*}

Fig.~\ref{fig:channel_maps} shows the channel maps of $^{12}$CO of PDS\,111 from ALMA observations

\end{appendix}

\end{document}